\documentclass[twocolumn,showpacs,preprintnumbers,amsmath,amssymb,superscriptaddress,groupedaddress,prmater]{revtex4-1}

\usepackage{epsfig}                                                            
\usepackage{graphicx}
\usepackage{dcolumn}
\usepackage{color}
\usepackage{bm}
\usepackage{subfigure}

\begin{document}

\title{\bf Electric-field-driven resistive transition in multiferroic SrCo$_2$Fe$_{16}$O$_{27}$/Sr$_3$Co$_2$Fe$_{24}$O$_{41}$composite}

\author {Shubhankar Mishra} \affiliation {School of Materials Science and Nanotechnology, Jadavpur University, Kolkata 700032, India}
\author {Aditi Sahoo} \affiliation {Advanced Mechanical and Materials Characterization Division, CSIR-Central Glass and Ceramic Research Institute, Kolkata 700032, India}
\author {Suchanda Mondal} \affiliation {Condensed Matter Physics Division, Saha Institute of Nuclear Physics, 1/AF Salt Lake, Kolkata 700064, India}
\author {P. Mandal} \email{Current address: S.N. Bose National Centre for Basic Sciences, Kolkata 700098, India} \affiliation {Condensed Matter Physics Division, Saha Institute of Nuclear Physics, 1/AF Salt Lake, Kolkata 700064, India} 
\author {Chandan K. Ghosh} \email {chandu\textunderscore ju@yahoo.co.in} \affiliation {School of Materials Science and Nanotechnology, Jadavpur University, Kolkata 700032, India}
\author {Dipten Bhattacharya} \email {dipten@cgcri.res.in} \affiliation {Advanced Mechanical and Materials Characterization Division, CSIR-Central Glass and Ceramic Research Institute, Kolkata 700032, India}

\date{\today}

\begin{abstract}
We report observation of electric-field-driven resistive transition [abrupt rise in resistivity ($\rho$)] at a characteristic threshold field $E_{th}(T)$ across a temperature range 10-200 K in an off-stoichiometric composite of W- and Z-type hexaferrite ($\sim$80\%)SrCo$_2$Fe$_{16}$O$_{27}$/($\sim$20\%)Sr$_3$Co$_2$Fe$_{24}$O$_{41}$. The dielectric constant $\epsilon$ and the relaxation time constant $\tau$ too exhibit anomalous jump at $E_{th}(T)$. The $E_{th}(T)$, the extent of jump in resistivity ($\Delta\rho$), and the hysteresis associated with the jump [$\Delta E_{th}(T)$] are found to decrease systematically with the increase in temperature ($T$). They also depend on the extent of non-stoichiometry. Several temperature-driven phase transitions have also been noticed both in the low and high resistive states (LRS and HRS). Application of magnetic field ($\sim$10-20 kOe), however, appears to have weaker influence on $\rho(T)$, $E_{th}(T)$, $\Delta\rho$, and $\Delta E_{th}(T)$ in the low temperature phase while reasonably significant influence in the phases stabilized at higher temperature. The temperature-driven conduction turns out to be governed by activated hopping of small polarons at all the phases with electric ($E$) and magnetic ($H$) field dependent activation energy $U(E,H)$. Interestingly, as the temperature is raised, the $E$-driven conduction at a fixed temperature evolves from $\textit{Ohmic}$ to $\textit{non-Ohmic}$ across 10-200 K and within 110-200 K, $\rho$ follows three-dimensional variable range hopping (3D-VRH) with stretched exponential $\sim$ $exp[(E_0/E)^4]$ or power law $\sim$ $(E_0/E)^m$ ($m$ varies within $\sim$0.6-0.7 and $\sim$0.6-0.8 at LRS and HRS, respectively) dependence depending on the localization length ($\zeta_E$) to diffusion length ($d_E$) ratio associated with $E$-driven conduction. The $\rho(E,T)$ turns out to be following universal scaling only at LRS within 10-110 K but not at higher temperature or at HRS. The entire set of observations starting from $E$-driven resistive jump to $T$-driven phase transitions and failure of the universal scaling law to adequately describe the observed $\rho(E,T)$ patterns across the entire $E-T$ range have been discussed within the framework of structural evolution of the point-defect (cation vacancies or oxygen excess) network and its influence on electronic conduction. The magnetocapacitive effect, measured under $\sim$20 kOe field, turns out to be substantial ($\sim$4\%-12\%) and exhibits clear anomaly at $E_{th}$. This comprehensive map of esoteric $\rho-E-T-H$ and $\epsilon-E-T-H$ patterns provides insights on defect driven effects in a composite useful for tuning both the resistive transition and multiferroicity.         
\end{abstract}
\maketitle

\begin{figure*}[ht!]
\begin{center}
   \subfigure[]{\includegraphics[scale=0.30]{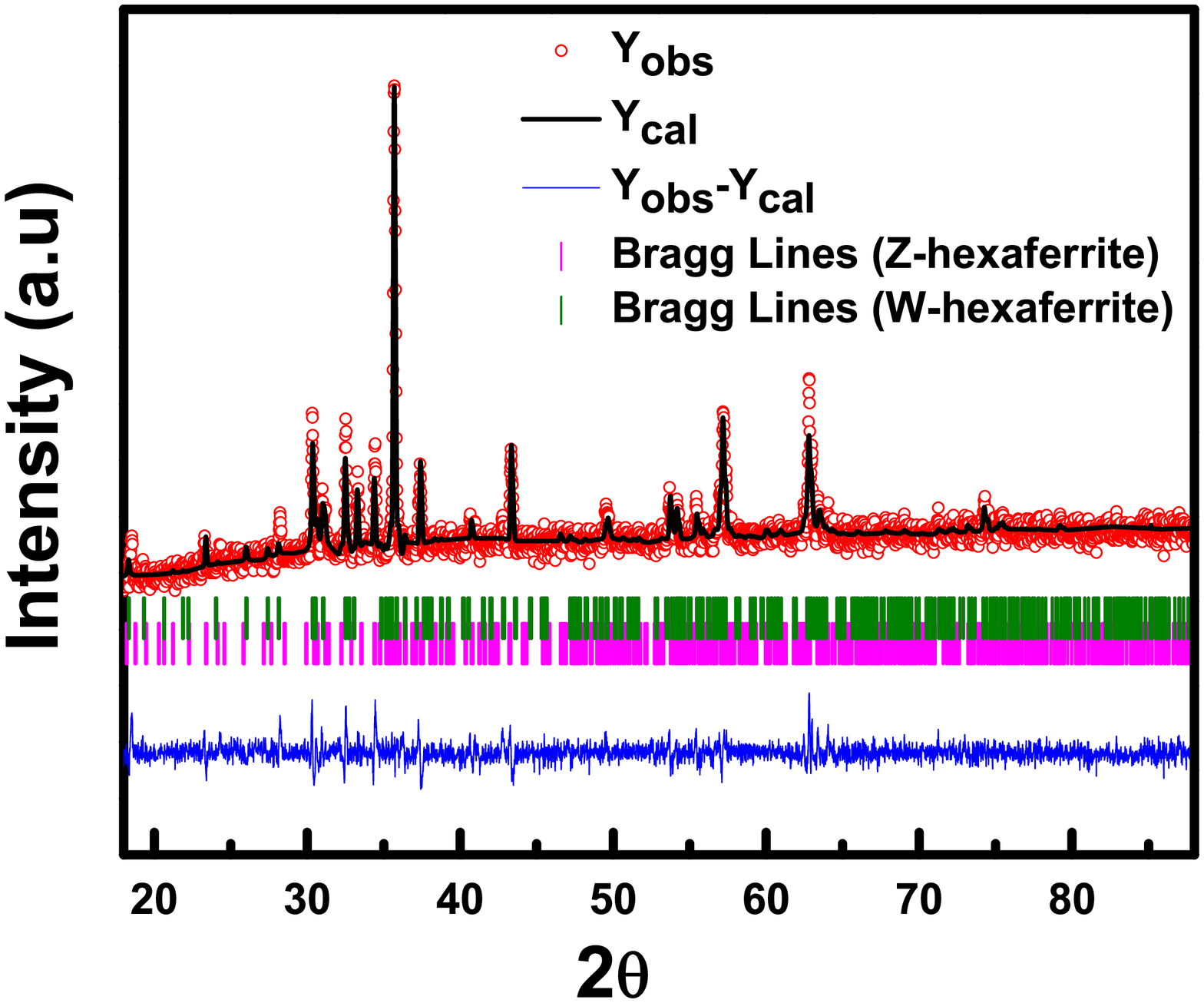}} 
   \subfigure[]{\includegraphics[scale=1.90]{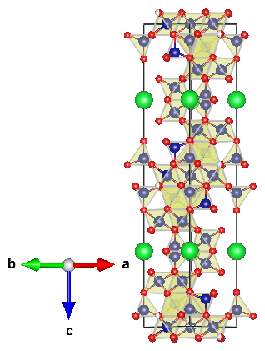}}
   \subfigure[]{\includegraphics[scale=2.20]{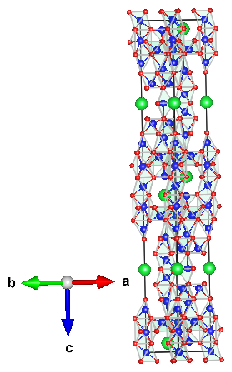}} 
   \subfigure[]{\includegraphics[scale=0.20]{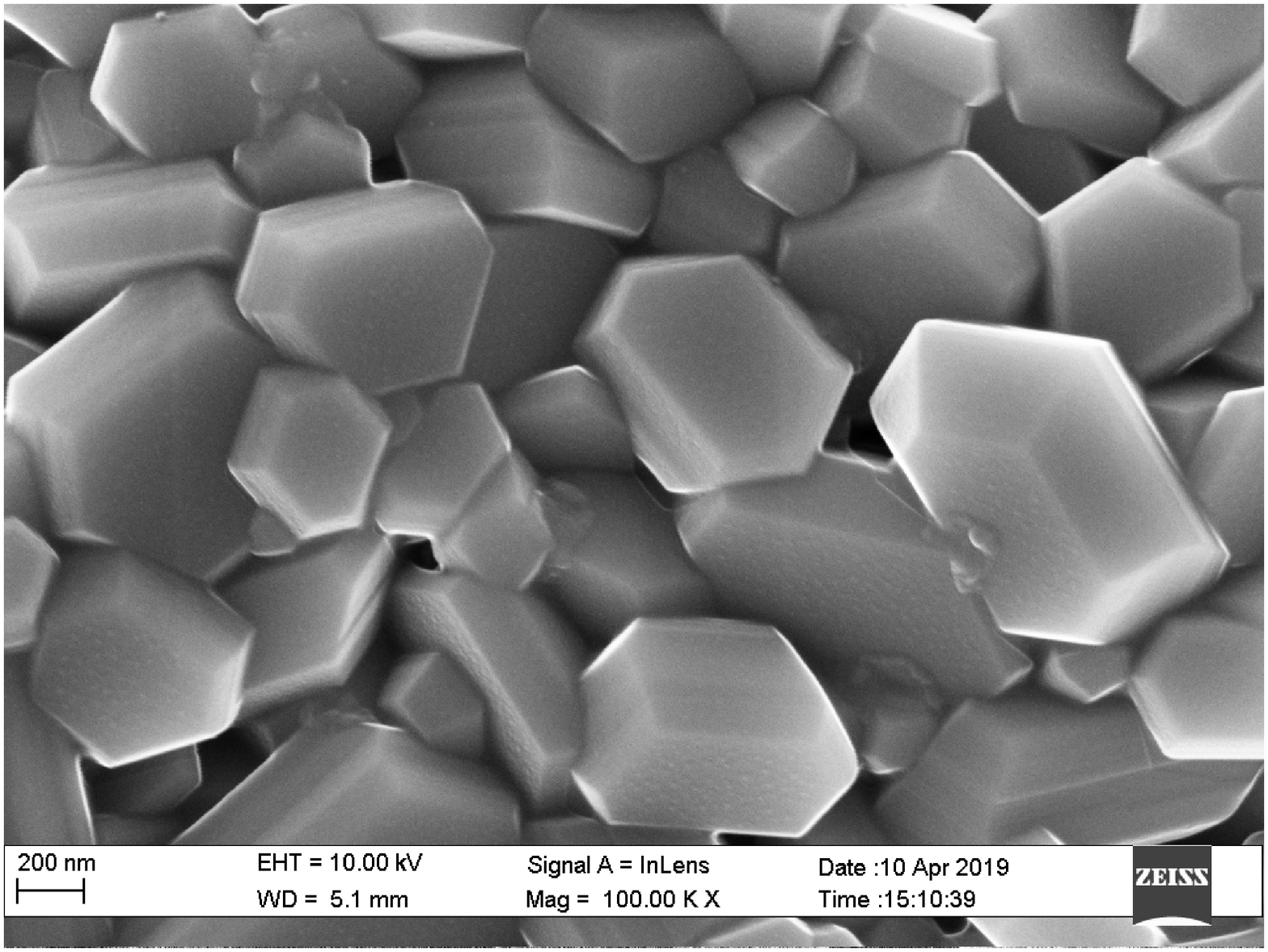}}  
   \end{center}
\caption{(a) The room temperature x-ray diffraction data and refinement by FullProf; (b) and (c) the crystallographic structure of W- and Z-type hexaferrites as obtained from the refinement; green, blue, and red spheres, respectively, represent Sr, Fe, and O; Co ions are not visible in these views; (d) the scanning electron microscopy image of the grain and grain boundaries in the sintered pellet. }
\end{figure*}

\section{Introduction}
The strongly correlated electron systems, because of subtle interplay among different degrees of freedom of electrons - charge, spin, orbital - and underlying lattice, exhibit stabilization of rather esoteric electronic, magnetic, and crystallographic phases \cite{Armitage}. In different temperature, magnetic field, mechanical pressure, composition, electric and/or optical field regimes, this subtle interplay gives rise to a variety of continuum or even granular phases, e.g., stripes, layered, one-dimensional, segregated etc \cite{Dagotto}. Study of these phases and phase transitions (both thermodynamic and kinetic) is rewarding not only for the rich underlying physics but also for opening vistas for a range of novel electronic/spintronic devices. Electric or optical pulse driven phase transitions such as melting of charge/orbital order, magnetic order, or crystallographic structure has already been observed \cite{Campi,Poccia,Li,Hassan} in a plethora of transition metal ion based strongly correlated compounds. 

It is also important to note that alongside such intrinsic phases and their transitions, presence of point defects too could give rise to new phases and exhibit external stimulation driven phase transitions \cite{Kalinin}. For example, resistance switching has been observed \cite{Kim} in many transition metal oxide systems because of electromigration of point defects - cation and/or oxygen vacancies. Electric field driven change in the point defect structure - formation and rupture of metallic filament - has been shown to be useful for developing resistive random access memories (RRAM) \cite{Zahoor}. Role of Joule heating in formation/rupture of metallic filaments has also been noted. Temperature driven order-disorder transition of point defects and consequent change in resistivity too, has been observed \cite{Tomura}. And, in general, presence of defects induces broadening of first order transition - classic examples of which are `driven vortex lattice transitions' observed in cuprate superconductors \cite{Vinokur}. 

Given this backdrop, we discover in a non-stoichiometric composite of W- and Z-type hexaferrites (containing W- and Z-type phaes by $\sim$80 and $\sim$20 vol\%) that application of electric field gives rise to a sharp rise in resistivity ($\rho$) beyond a certain threshold field. The threshold field $E_{th}$, the extent of jump in resistivity $\Delta\rho$, and the hysteresis associated with the transition $\Delta E_{th}$ are found to depend significantly on temperature ($T$). Within the regime 10-200 K, they decrease systematically with the rise in temperature. This abrupt rise in $\rho$ signifies order-disorder transition of the cation/oxygen vacancies under the application of electric field. The intrinsic dielectric constant - measured in presence of a dc bias - too, is found to exhibit anomalous features at the threshold field. Simultaneous application of magnetic field has weaker influence in the low temperature regime (10-25 K) but relatively stronger one in the regime $>$25 K. Interestingly, variation of the extent of non-stoichiometry is found to influence the electric field driven phase transition features such as $E_{th}$, $\Delta\rho$, and $\Delta E_{th}$ significantly. This observation underscores the possibility of tuning the electric field driven transition of ordered point defect phases within the host matrix via variation in the point defect concentration.

The Z-type hexaferrites are potential room temperature multiferroic compounds where, because of spiral magnetic structure, lattice noncentrosymmetry (and hence ferroelectric polarization) sets in under as small a magnetic field as $\sim$300 Oe \cite{Kitagawa,Soda}. The magnetic structure changes to simple ferrimagnetic one beyond $\sim$30 kOe with the disappearance of ferroelectricity. Since the first report on its room temperature multiferroicity, several studies have been carried out to elucidate the crystallographic, magnetic, and electronic structures. The crystallographic structure of hexaferrites assumes $P6_3/mmc$ symmetry and is composed of several blocks designated as R, S, and T depending on their close packing of oxygen layers. In the case of Z-type ones, the blocks are rotated by 180$^o$ around c-axis; the rotated ones are designated by asterisks such as R$^*$, S$^*$, T$^*$. The Sr$_3$Co$_2$Fe$_{24}$O$_{41}$ compound is composed of stacking of RSTSR$^*$S$^*$T$^*$S$^*$ blocks. Several magnetic transitions take place over a temperature range 300-700 K. The paramagnetic to ferrimagnetic transition takes place at $T_N$ $\approx$ 670 K (with moments aligned along c-axis) while below $\sim$500 K, the moments reorient away from c-axis by $\sim$50.5$^o$. Finally, below $\sim$400 K, transverse conical spin structure emerges which breaks the $P6_3/mmc$ symmetry. This structure, of course, systematically changes under the application of magnetic field and beyond $\sim$30 kOe, it becomes ordinary ferrimagnetic again. Ferroelectric polarization appears and peaks (at $\sim$ 3 nC/cm$^2$) near $\sim$300 Oe and decreases thereafter. 

The W-type hexaferrite SrCo$_2$Fe$_{16}$O$_{27}$, on the other hand, is composed of RSSRS$^*$S$^*$ blocks and assumes $P6_3/mmc$ symmetry \cite{Morch}. It exhibits ferromagnetic order below $T_C$ $\sim$ 850 K and below $T^*$ $\sim$550 K, conical magnetic structure sets in. However, it does not exhibit finite ferroelectric polarization - neither due to structural instability nor because of symmetry breaking magnetic structure. Below 300 K and down to 10 K, no further magnetic transition could be noticed in either W- or Z-type hexaferrite.

At room temperature, the W-type system is, therefore, primarily magnetic while the Z-type one exhibits ferroelectricity under a magnetic field. We endeavor to develop a series of composites of W- and Z-type hexaferrites covering the entire range from pure W-type to pure Z-type to map out the evolution of multiferroicity within such composite systems. However, very interestingly, we discovered that non-stoichiometry in a composite gives rise to formation of ordered point defect networks which, in turn, exhibit remarkable array of electric field and temperarure driven phases and phase transitions. In this paper, therefore, we focus on the point defect phases and their transitions which appear to influence the magnetoelectric properties of the composite as well.

\begin{figure*}[ht!]
\begin{center}
   \subfigure[]{\includegraphics[scale=0.25]{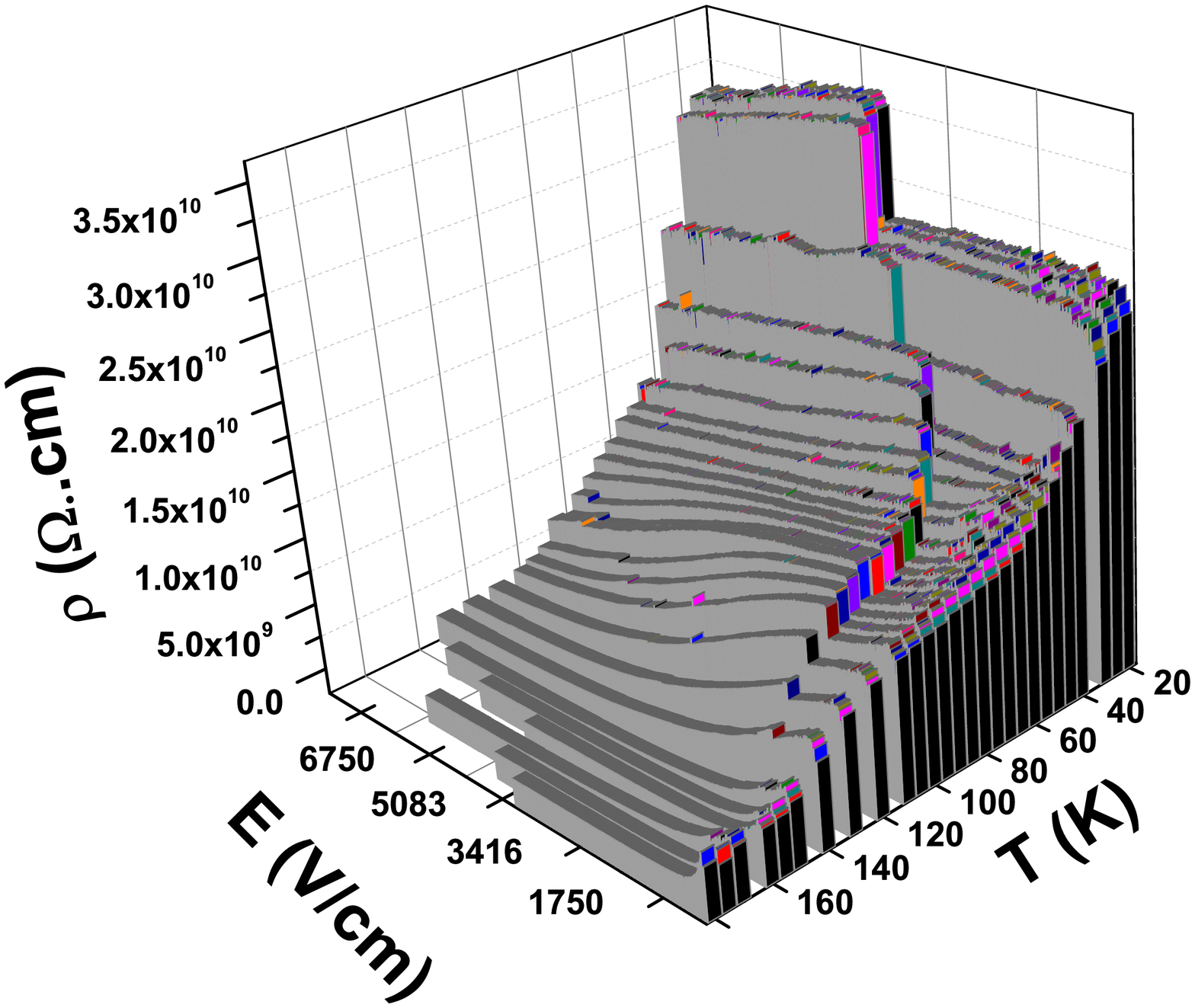}}
   \subfigure[]{\includegraphics[scale=0.15]{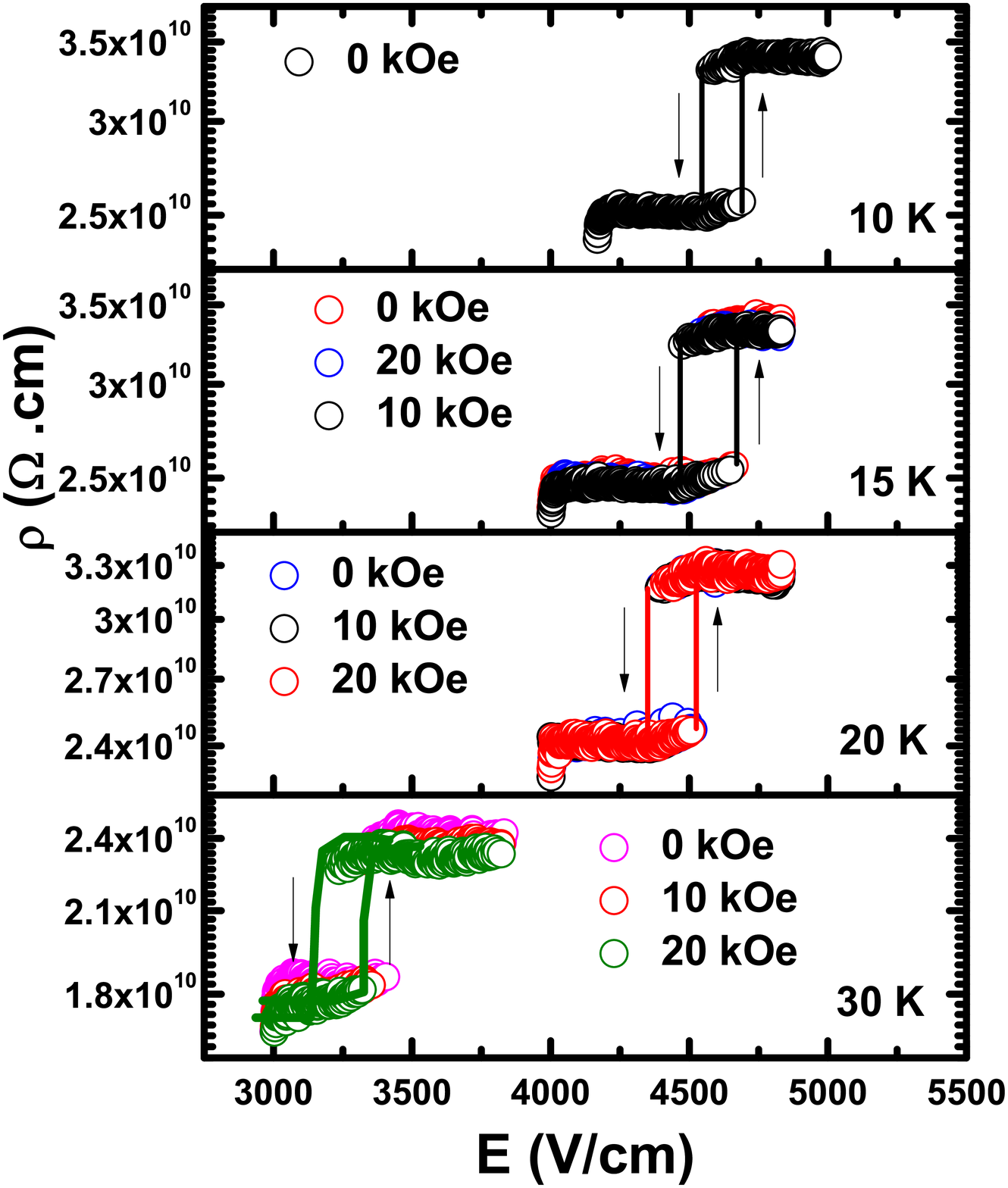}} 
   \subfigure[]{\includegraphics[scale=0.15]{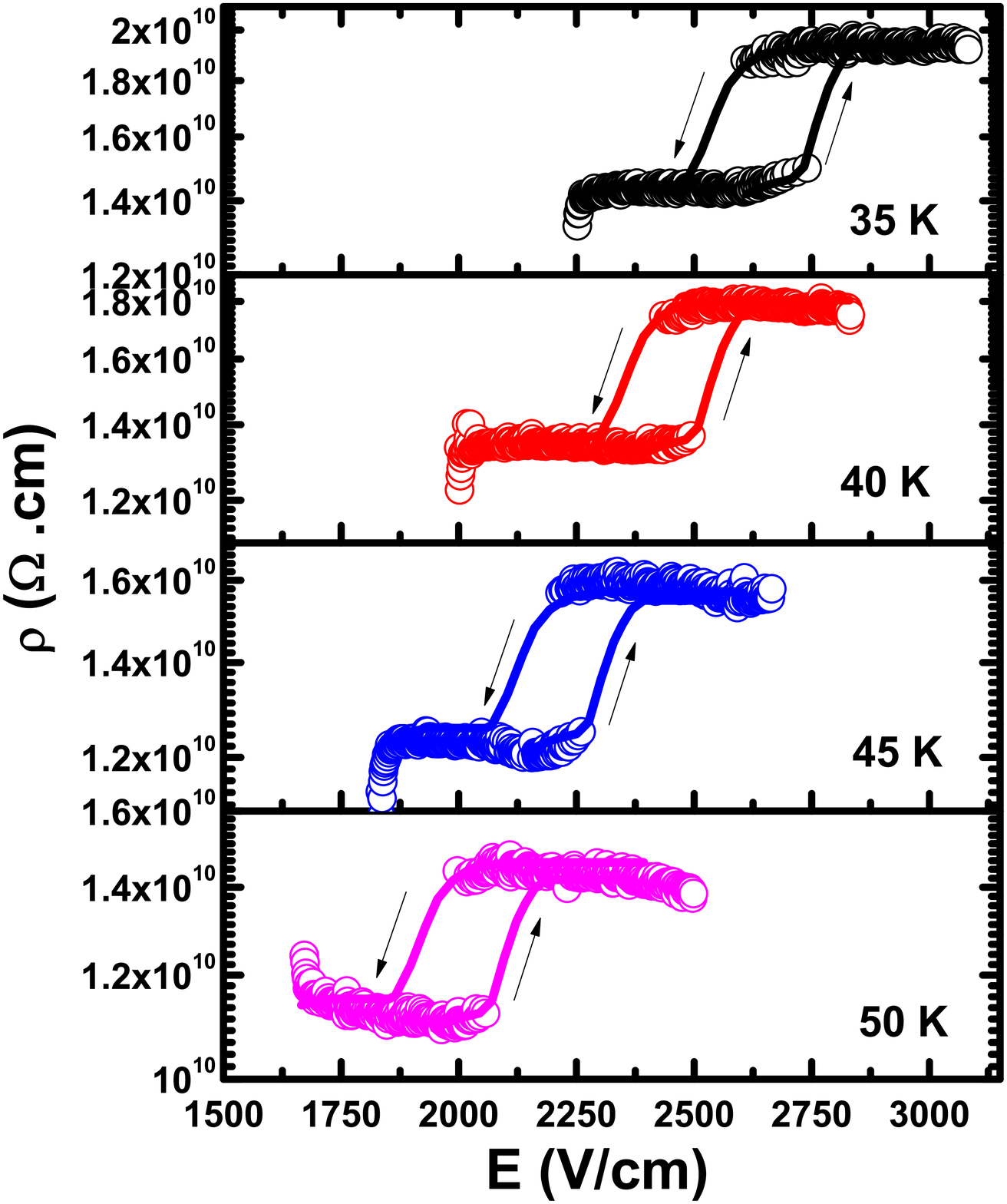}}
   \subfigure[]{\includegraphics[scale=0.15]{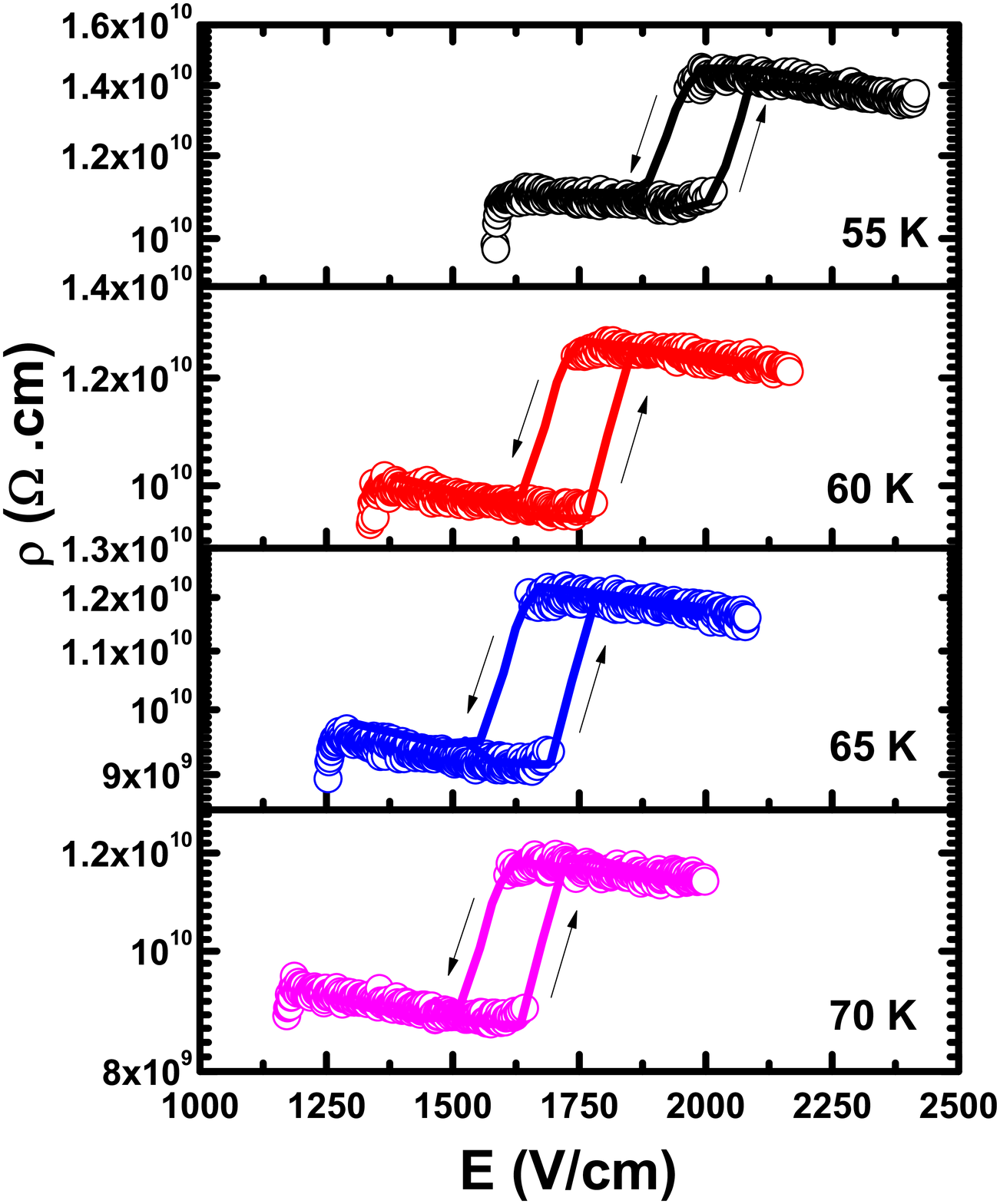}} 
   \subfigure[]{\includegraphics[scale=0.15]{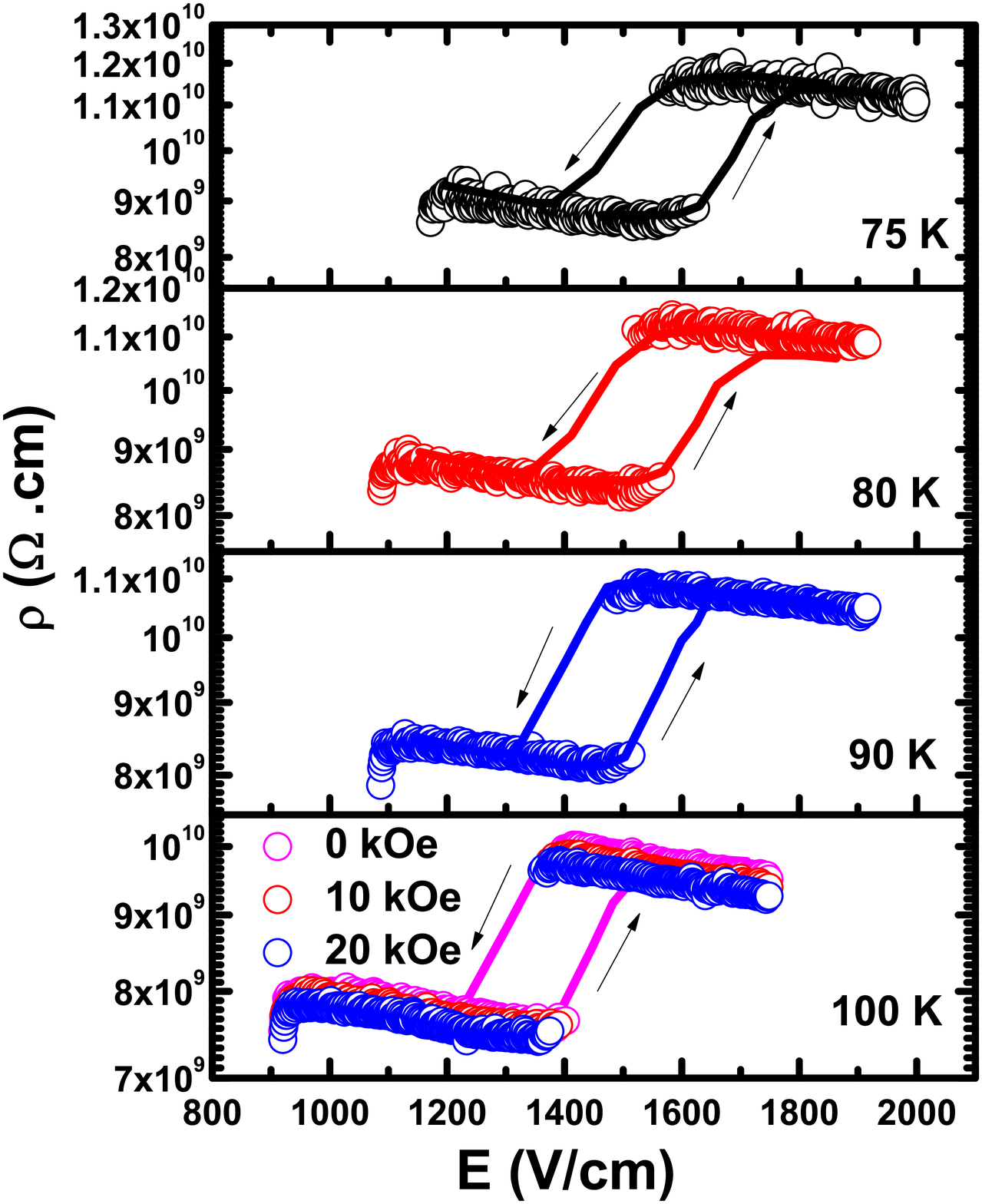}} 
   \subfigure[]{\includegraphics[scale=0.15]{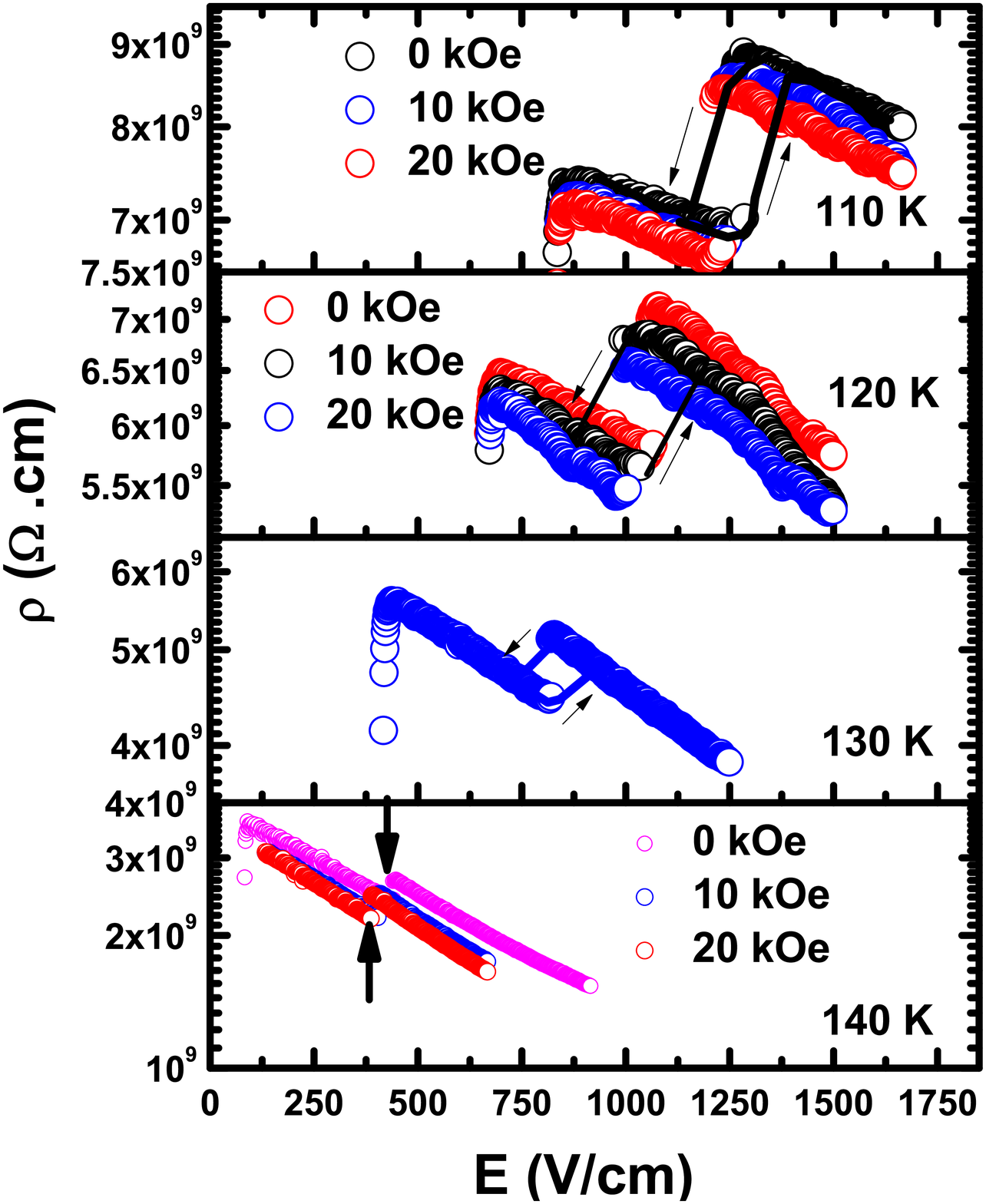}}
   \subfigure[]{\includegraphics[scale=0.15]{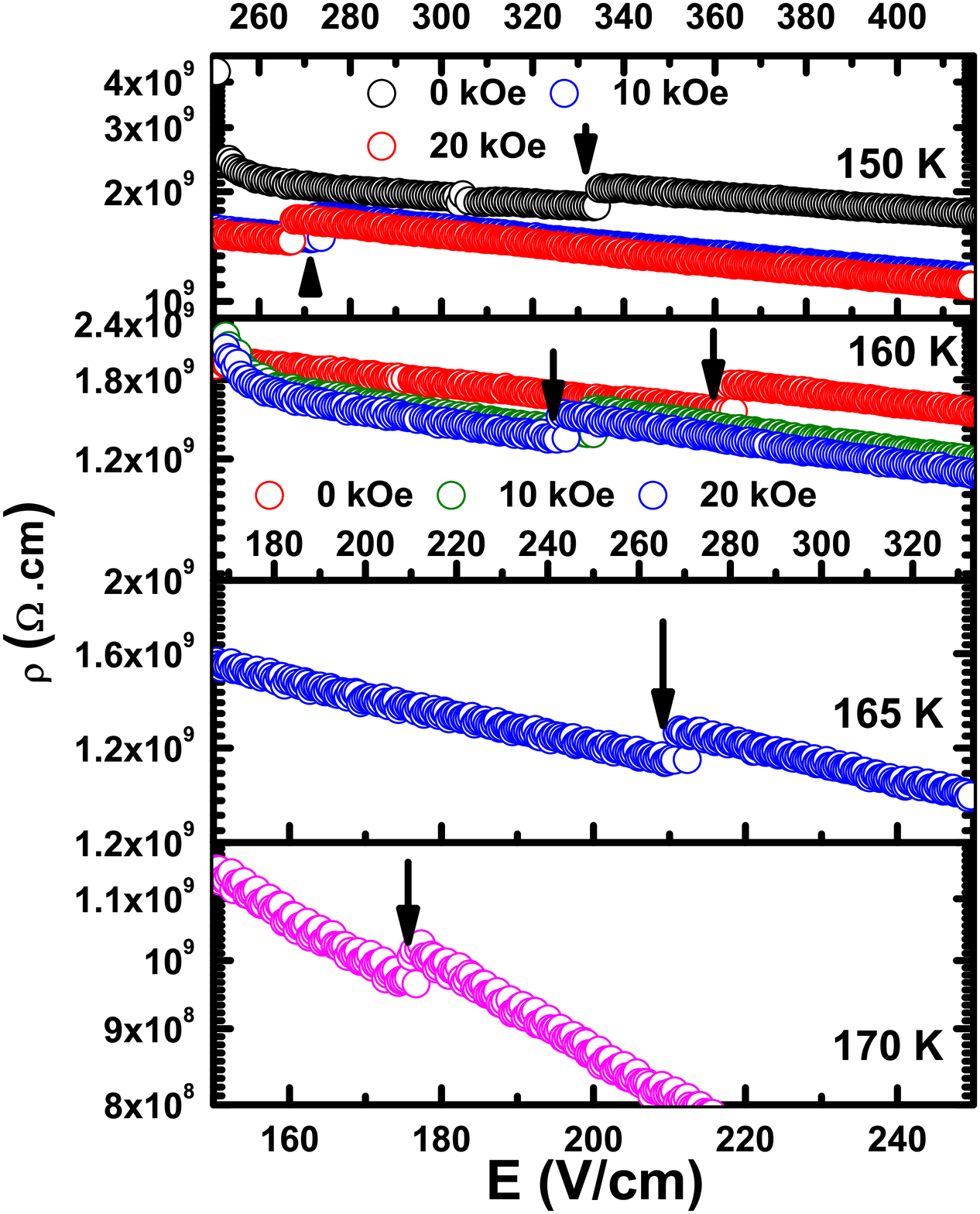}}
   \subfigure[]{\includegraphics[scale=0.20]{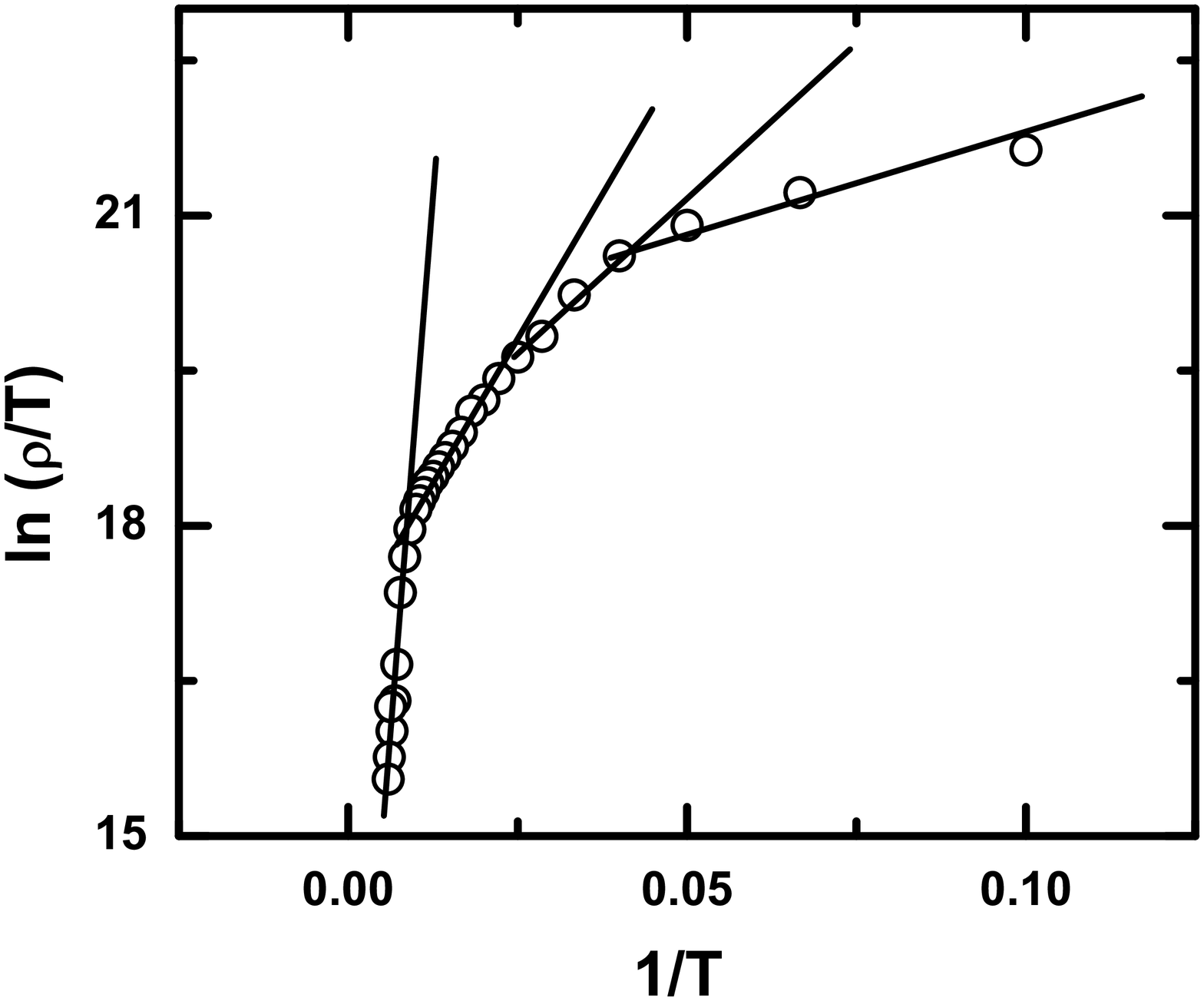}} 
   \subfigure[]{\includegraphics[scale=0.20]{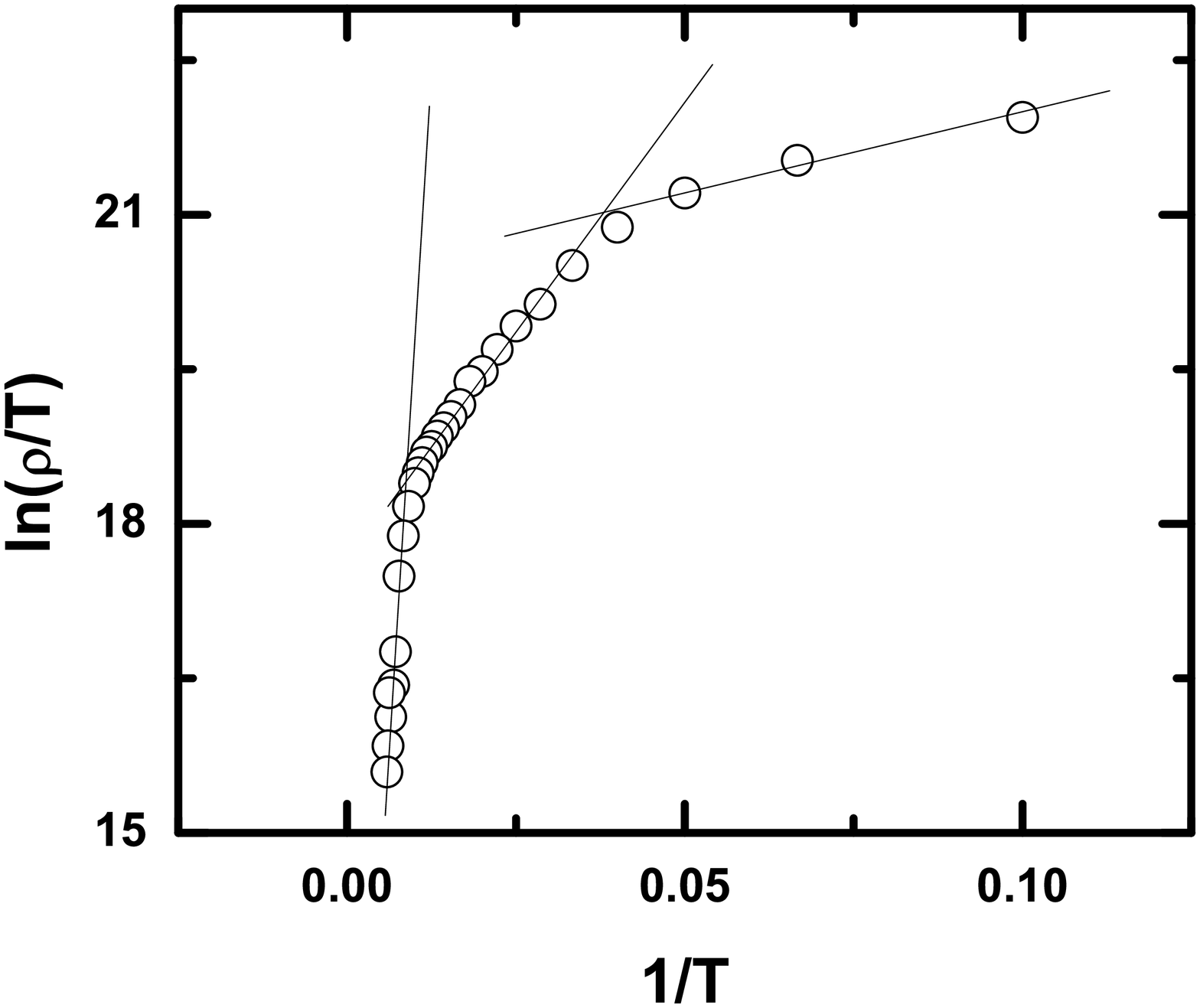}}  
   \end{center}
\caption{(a) The three-dimensional resistivity-temperature-electric field patterns; (b-g) the two-dimensional resistivity versus electric field patterns in different temperature regimes; the lines and arrows highlight the hysteresis in the transition; ln($\rho/T$) versus 1/$T$ plot at (h) LRS and (i) HRS under zero magnetic field. }
\end{figure*}

\begin{figure*}[ht!]
\begin{center}
   \subfigure[]{\includegraphics[scale=0.15]{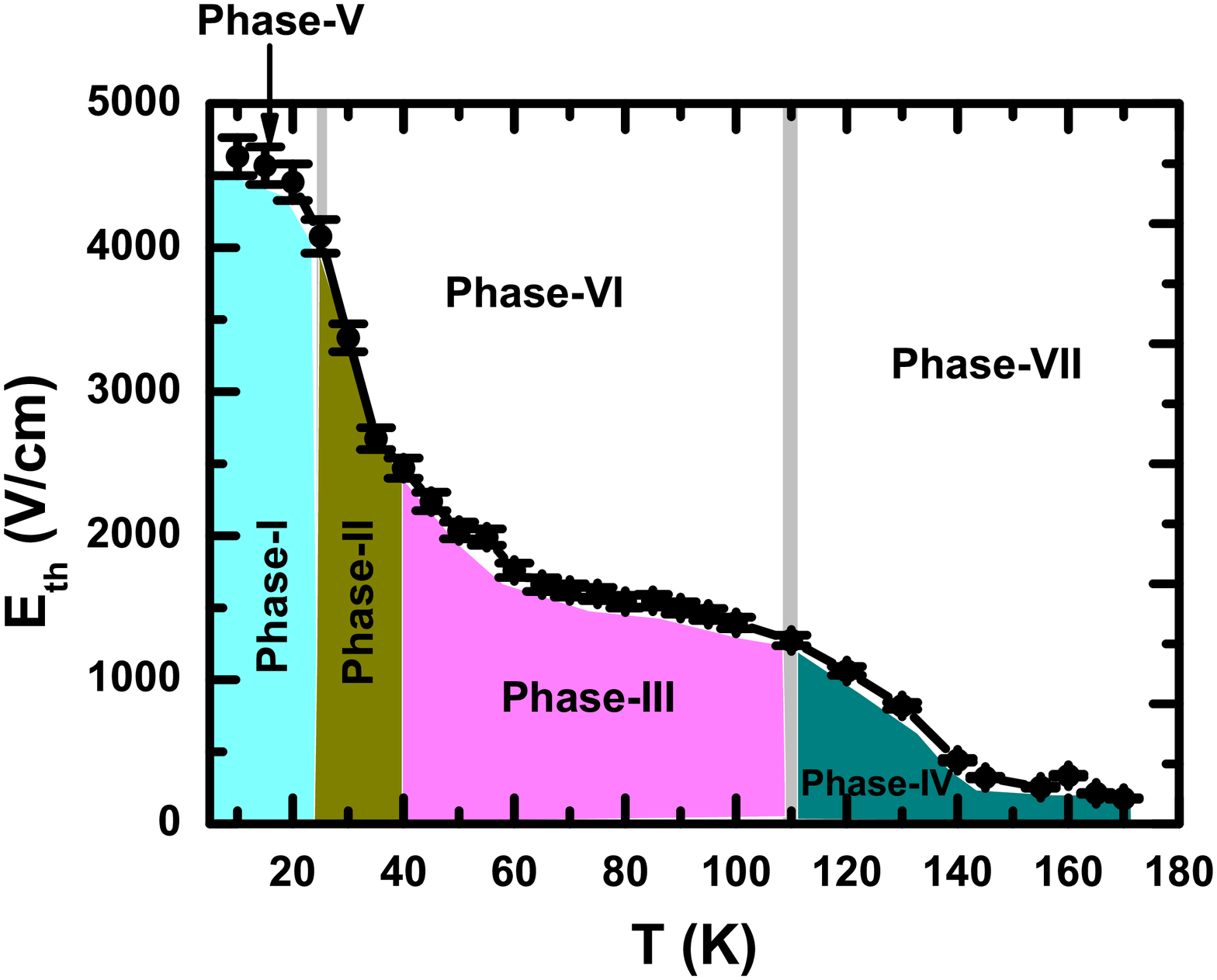}} 
   \subfigure[]{\includegraphics[scale=0.15]{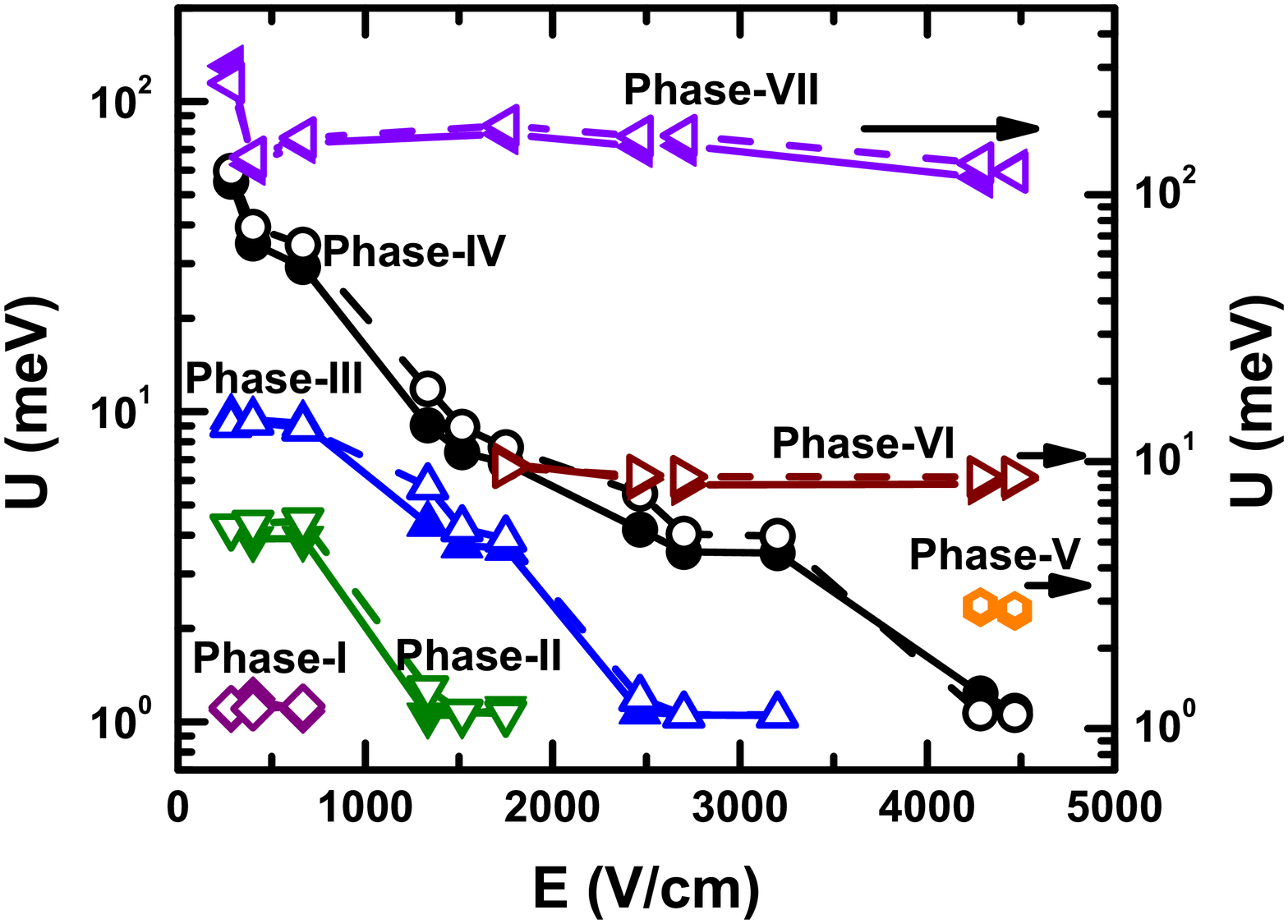}}
   \subfigure[]{\includegraphics[scale=0.15]{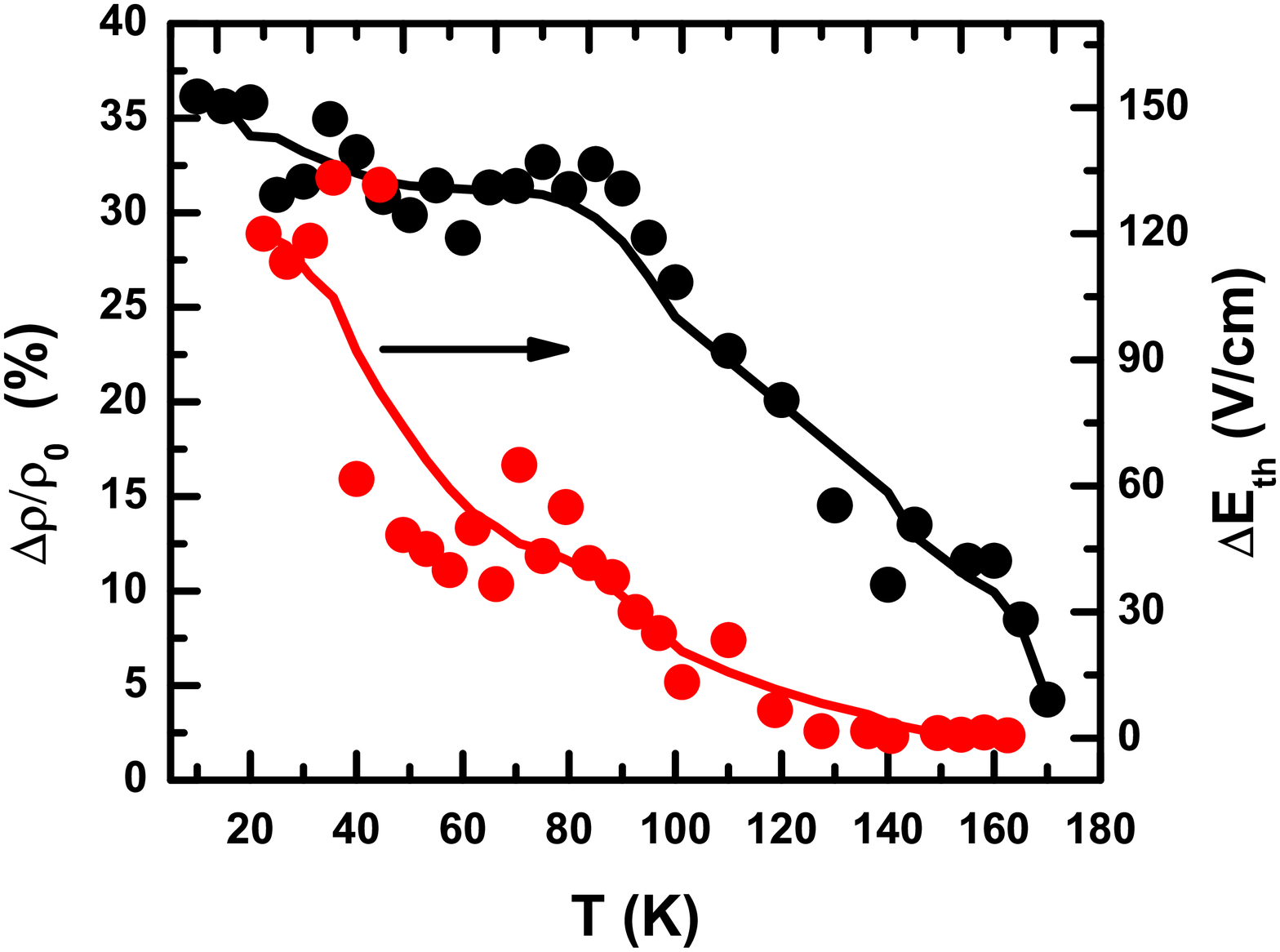}}
   \subfigure[]{\includegraphics[scale=0.15]{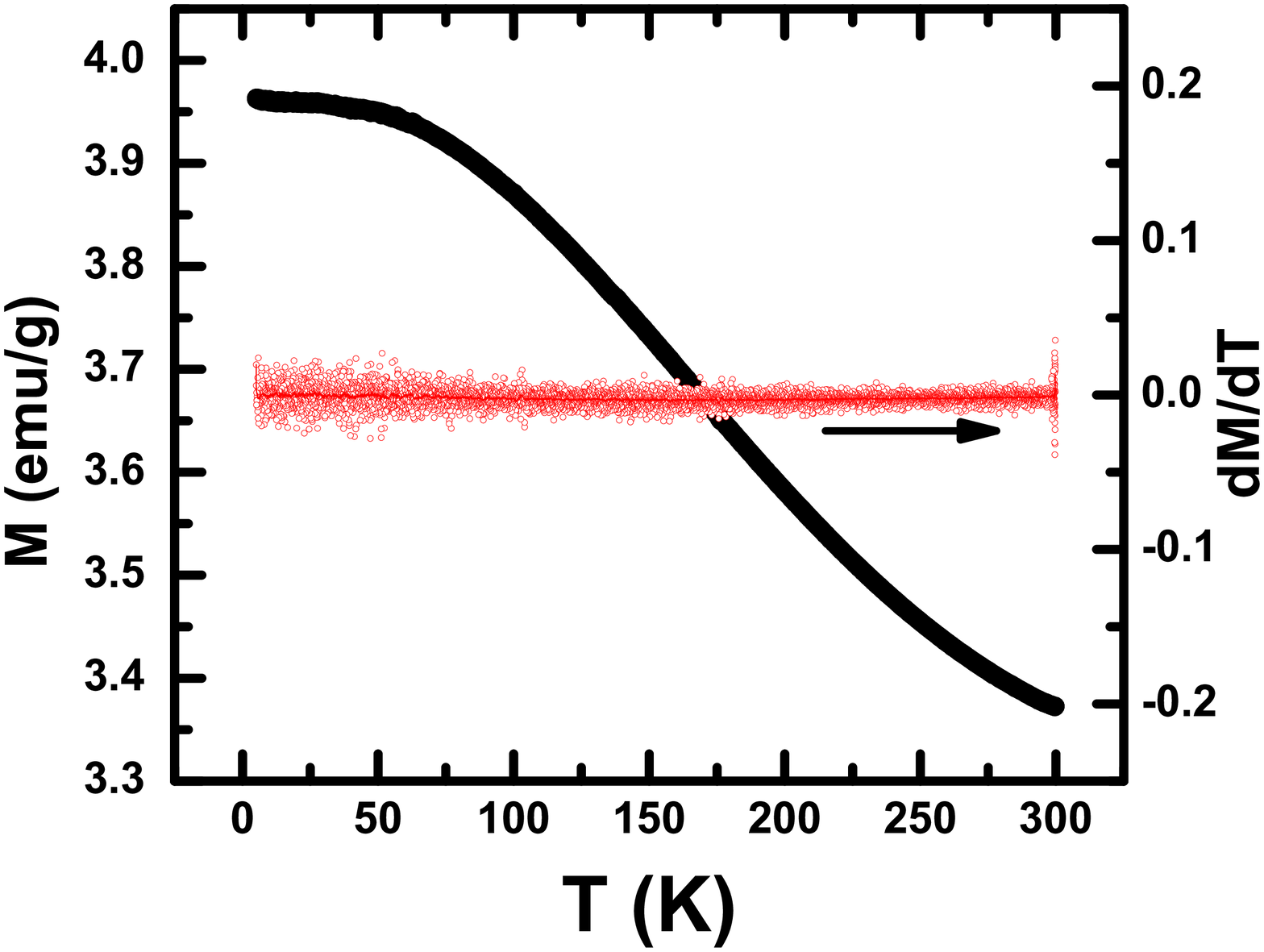}}
   \end{center}
\caption{(a) The $E_{th}-T$ phase diagram; (b) variation of $U$ (solid symbols) with $E$ across different phases identified in (a); influence of simultaneous application of $H$ = 20 kOe on $U$ (open symbols) is discernible only in phase-IV and phase-VII (c) variation of $\Delta \rho$/$\rho_0$ and $\Delta E_{th}$ with temperature; (d) field-cooled (FC) magnetization ($M$) versus temperature ($T$) and $dM/dT$ versus $T$ data across 5-300 K; as expected, no magnetic transition could be noticed in this temperature range. }
\end{figure*}

\begin{figure}[ht!]
\begin{center}
   \subfigure[]{\includegraphics[scale=0.20]{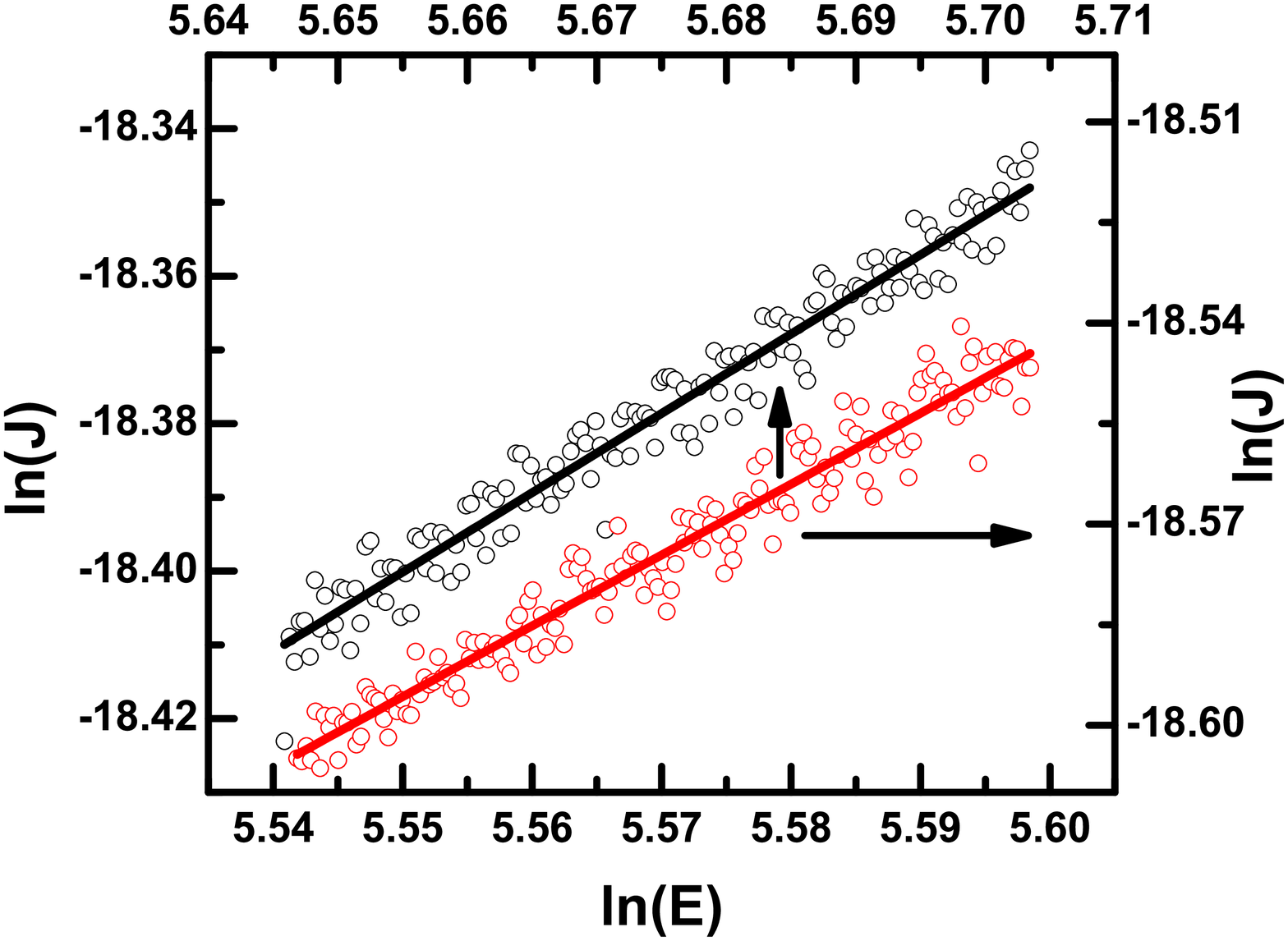}} 
   \subfigure[]{\includegraphics[scale=0.20]{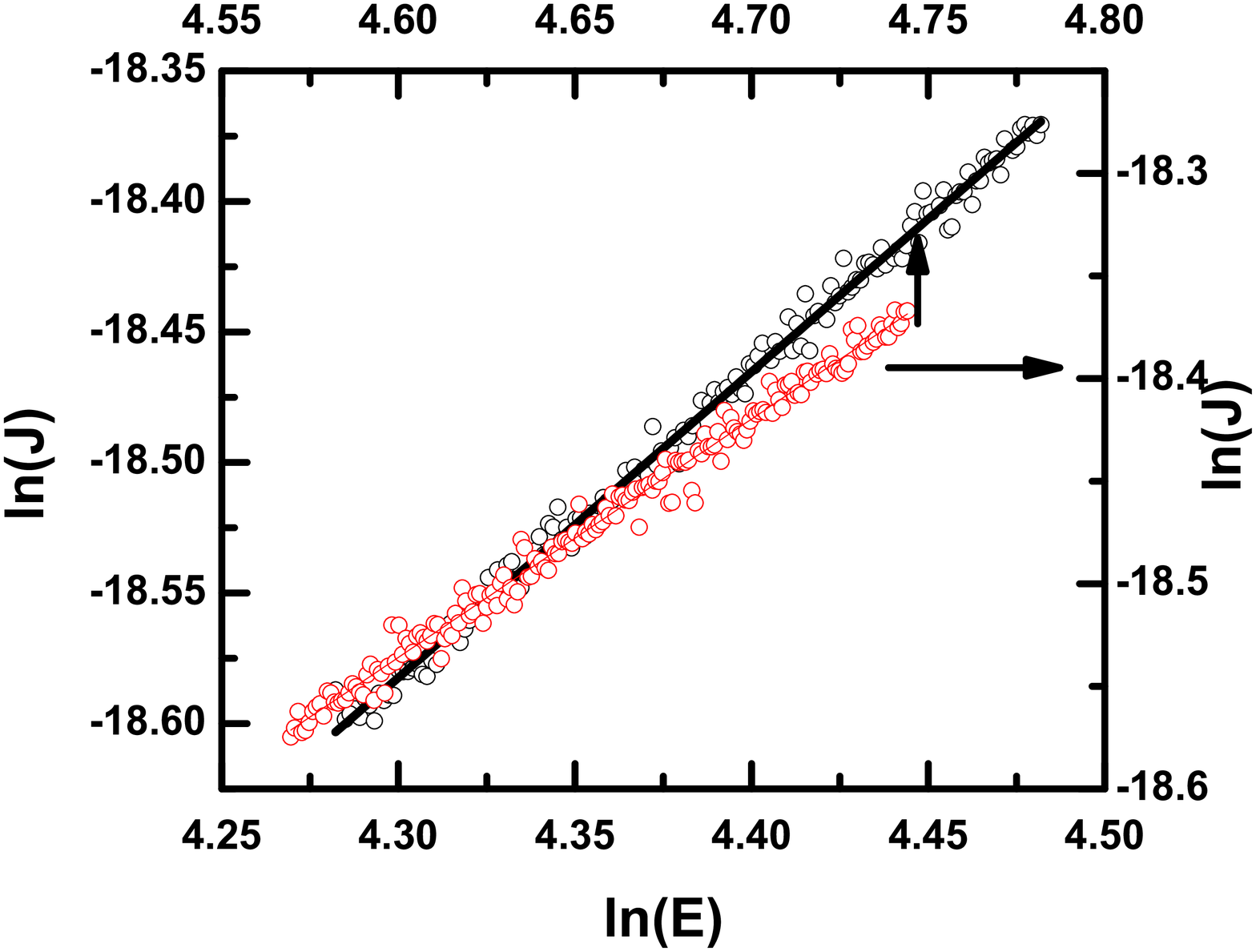}}
   \subfigure[]{\includegraphics[scale=0.20]{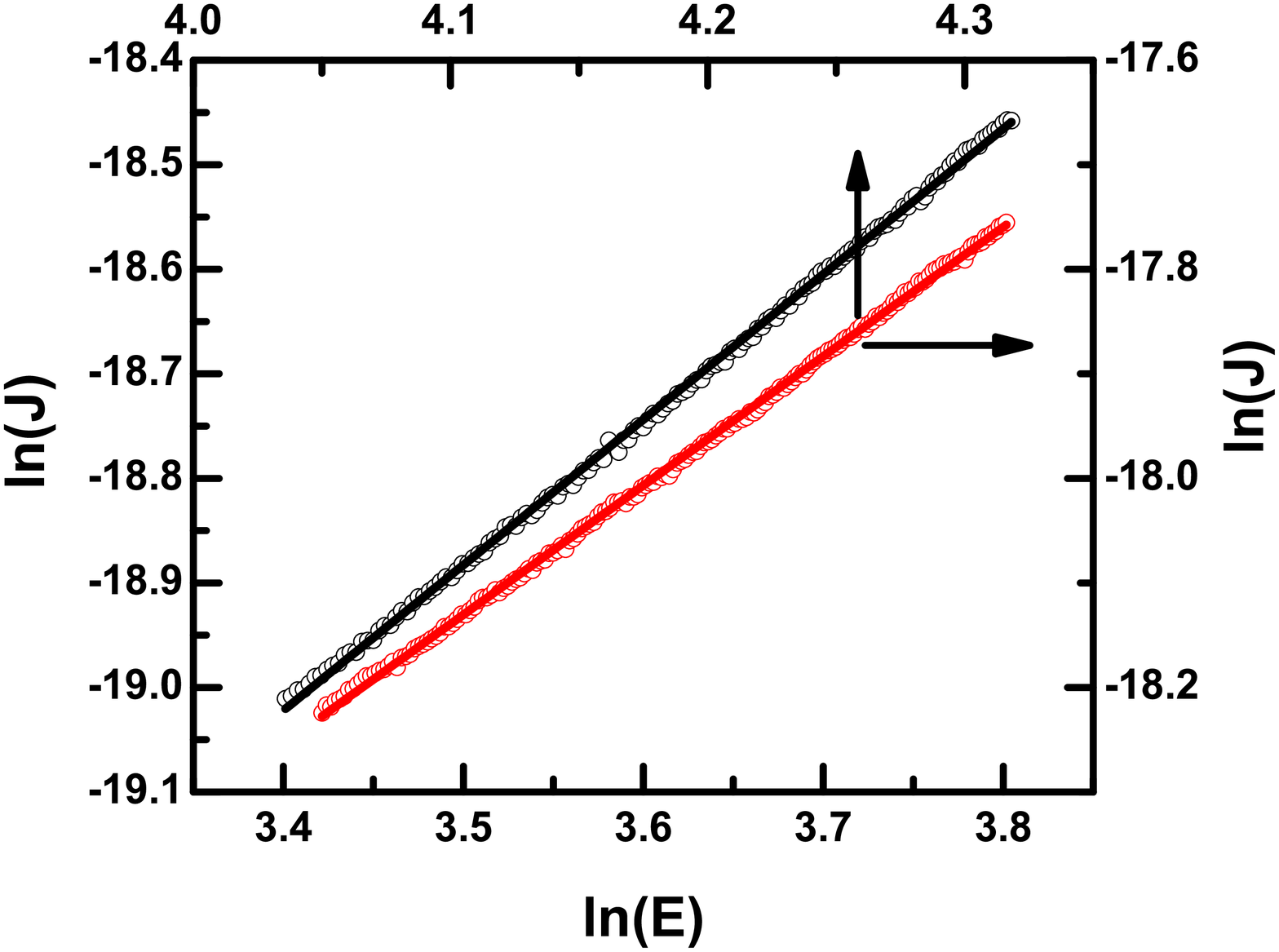}}
   \end{center}
\caption{The $J-E$ patterns at LRS (black symbols) and HRS (red symbols) in (a) low temperature (10 K), (b) intermediate temperature (80 K) and (c) high temperature (140 K) regimes. }
\end{figure}

\section{Experimental}
The bulk polycrystalline samples were prepared by solid state reaction technique using high purity ($>$99.9\%) carbonates and oxides SrCO$_3$, Co$_3$O$_4$, and Fe$_2$O$_3$. They were mixed in stoichiometric proportion (based on nominal compositions) using agate balls in a ball mill under ethanol medium. The mixed and dried powder was subsequently calcined in air at 1100$^o$C for 16h. The calcined powder was finally compacted in the form of pellets and sintered at 1200$^o$C for 16h in air. In order to eliminate the impurity phases, the samples were quenched to room temperature. The extent of non-stoichiometry was varied by treating the samples at different temperature ranging from 1200$^o$C to 1400$^o$C and also under flowing oxygen. Although a series of composites containing different volume fractions of W- and Z-type hexaferrites has been prepared, for this work, we concentrate on the ones which contain the W- and Z-type systems by $\sim$80 and $\sim$20 vol\%, respectively, and are off-stoichiometric by different extent. The sintered pellets were characterized by x-ray diffraction (XRD), scanning electron microscopy and energy dispersive x-ray spectra (SEM and EDX), electrical, and magnetic measurements. The electrical resistivity of the samples was measured in two-probe configuration by using Keithley 6220 current source and HP3458A multimeter and HP4339B high resistance meter. Top and bottom silver electrodes were used on the pellet samples of diameter $\sim$5 mm and thickness $\sim$0.5 mm. The magnetic measurements were carried out by a vibrating sample magnetometer (VSM) of LakeShore (Model 7407) while the dielectric spectroscopy was recorded by an Impedance Analyzer (IM3536, Hioki).  
 
\section{Results and Discussion}
Figure 1a shows the room temperature x-ray data of the composite. The pattern was refined by FullProf to yield the structural details of each of the phases along with their volume fractions. As expected, the crystallographic structure assumes $P6_3/mmc$ symmetry for both the phases with lattice parameters $a$ = $b$ = 5.849, $c$ = 33.112 $\AA$ for W-type and $a$ = $b$ = 5.841, $c$ = 51.949 $\AA$ for Z-type systems. The multiphase fitting yields the volume fraction of the respective phases to be $\sim$80\% W-type and $\sim$20\% Z-type. The ion positions, bond lengths, and bond angles for each of the phases have been given in the supplementary document \cite{supplementary}. The structures, constructed by using the details obtained from x-ray diffraction data, are shown in Figs. 1b and 1c. Figure 1d shows a typical SEM image. The average grain size turns out to be $\sim$0.5 $\mu$m. The EDX spectra \cite{supplementary}, however, show the composite to be off-stoichiometric. Mapping of the concentration of the elements Sr, Co, Fe, and O at several different regions of the sample provides detailed information about the overall composition. Quantitative estimation of the atomic concentration of the elements yields the average actual composition and thereby indicates the extent of point defects \cite{supplementary}. The phases, therefore, contain sizable cation vacancies as well as oxygen excess. The EDX spectra corresponding to the samples containing different extent of off-stoichiometry are also shown in the supplementary document \cite{supplementary}. Yet, within this range of non-stoichiometry, no obvious change in crystallographic structure could be noticed for any of the phases in laboratory powder x-ray diffraction experiment. 

We carried out measurements of current-voltage characteristics across 10-300 K. The voltage is swept from zero to positive maximum and then from positive maximum to zero. Corresponding current is measured at each step. Typically, 1000 steps have been used in between zero and maximum applied field. Figure 2(a) summarizes the $\rho-E-T$ patterns in a three-dimensional plot while Figs. 2(b-g) present the blown-up features: dependence of $\rho$ on the applied electric field ($E$) at different temperatures. The corresponding $J-E$ patterns are shown in the supplementary document \cite{supplementary}. Following remarkable features could be observed in the entire set of the data. (i) At different temperatures over a range 10-200 K, $\rho(E)$ exhibits a sharp rise at a characteristic threshold field $E_{th}$. The transition from low to high resistive states (LRS and HRS) is reversible yet associated with a hysteresis ($\Delta E_{th}$). Using $E_{th}(T)$ data, a phase diagram could be constructed (Fig. 3a). It is important to mention here that unlike in the bipolar cases where application of positive bias gives rise to sharp change in resistivity and negative bias restores it, the electric field driven resistive transition observed here is unipolar in nature \cite{Waser}. The resistivity is restored as the applied field is withdrawn irrespective of the sign of the field. Joule heating alone cannot be responsible in inducing this transition as the transition is taking place from LRS to HRS at $E_{th}$. The raw $\rho-T$ data (Fig. 2h) corresponding to the LRS, of course, shows the expected drop in resistivity with the rise in temperature. (ii) Interestingly, $\rho-E$ patterns (at a fixed $T$) evolve from $Ohmic$ to $non-Ohmic$ as the temperature is raised. It follows stretched exponential or even power-law dependence in the higher temperature regime. (iii) The $\rho-T$ data at a fixed $E$ corresponding to LRS, on the other hand, also exhibit anomalous transition features at several temperatures $T_1$ $\approx$ 25 K, $T_2$ $\approx$ 40 K, and $T_3$ $\approx$ 110 K. It indicates further transitions below 300 K and prevalence of four different phases - phase-I, phase-II, phase-III, and phase-IV - in the 10-200 K range (Fig. 3a). The data corresponding to HRS exhibit transitions only at $T_1$ $\approx$ 25 K and $T_3$ $\approx$ 110 K. The $\rho-T$ patterns in all these regimes follow the adiabatic small polaron hopping mechanism where $\rho (T)$ is given by $\rho (T)$ = $\rho_0T$.exp($U/k_BT$) ($U$ = activation energy); $U$ turns out to be decreasing with the increase in $E$ in each of the phases of LRS and HRS. The $U$, corresponding to each of the phases (of LRS and HRS), as a function of $E$ has been mapped out in Fig. 3b. Distinct $U$ corresponding to a specific temperature zone clearly indicates prevalence of distinct phases and temperature driven phase transition. Figure 3c shows the variation of the $\Delta \rho$/$\rho_0$ and the extent of hysteresis in $E_{th}$ ($\Delta E_{th}$) with temperature ($T$). (iv) Application of magnetic field ($H$ = 10, 20 kOe) appears to have weaker influence in phase-I while reasonably stronger one in phases-II, III etc.

Similar measurements were carried out on different samples containing different extent of non-stoichiometry. Salient results have been included in the supplementary document \cite{supplementary} along with the raw $\rho-T$ data. Interestingly, all the important parameters such as $E_{th}$, $\Delta\rho$, $\Delta E$ etc exhibit dependence on the point defect concentration. For samples with smaller defect concentration, $E_{th}$ turns out to be higher while $\Delta\rho$ and $\Delta E$ are lower. The effect of electromigration is small. As the defect concentration enhances, the effect too rises and becomes significant. This could be understood by noting the chemical strain field generated within the host matrix by the point defects. For smaller defect concentration, the chemical strain is smaller. This, in turn, requires higher $E$ to induce structural transition in the defect network. With the rise in defect concentration the chemical strain field enhances which yields smaller $E_{th}$. Recently, interplay among the chemical strain field associated with point defects and strain generated due to electromigration and mechanical nanoprobing has been studied \cite{Yao} within an off-stoichiometric La$_{\frac{2}{3}}$Sr$_{\frac{1}{3}}$MnO$_3$ epitaxial thin film. Under simultaneous application of voltage pulse and mechanical force, the point defects were shown to undergo a structural transition from two- to three-dimensional ordering. Elastic strain originating from defects, electrical, and mechanical forces influences the structural transition of the point defect network. Complete mapping of this effect as a function of point defect concentration, of course, is beyond the scope of the present paper. We focus here on the sample which exhibits significant $E$-driven phase transition in order to unravel in greater detail the complexity surrounding the $E$-driven defect migration and its consequences. 

In order to trace the origin of the temperature-driven phase transitions, we carried out field-cooled (FC) magnetization ($M$) versus temperature ($T$) measurements across 5-300 K. The data are shown in Fig. 3d. We also plot the $dM/dT$ versus $T$ pattern in Fig. 3d to examine the signature of the magnetic transition. Expectedly, no magnetic transition could be noticed in this composite around the temperatures $T_1$, $T_2$, and $T_3$. This is because neither W-type nor Z-type hexaferrite is known to exhibit any magnetic transition in the 5-300 K range \cite{Soda,Morch}. The extent of off-stoichiometry - observed in the present case - turns out to be not sufficient to induce any magnetic transition within the 5-300 K range. The transitions observed in $\rho-T$, therefore, could arise from structural evolution of the point defect network but not from transitions in grain/domain boundary features. Clear observation of $T$-driven activated adiabatic small polaron hopping conduction implies bulk conduction.   

\begin{figure}[ht!]
\begin{center}
   \subfigure[]{\includegraphics[scale=0.20]{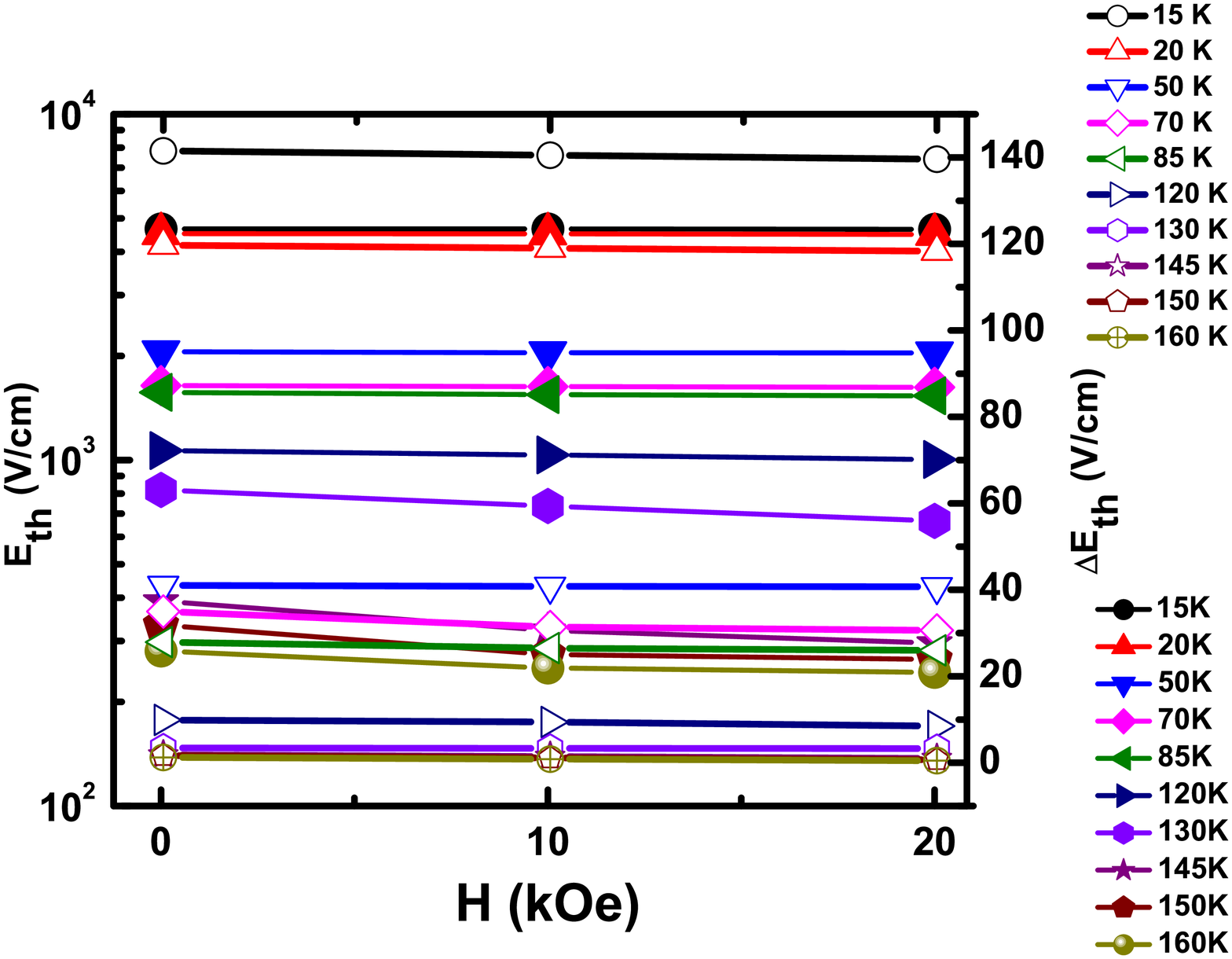}} 
   \subfigure[]{\includegraphics[scale=0.20]{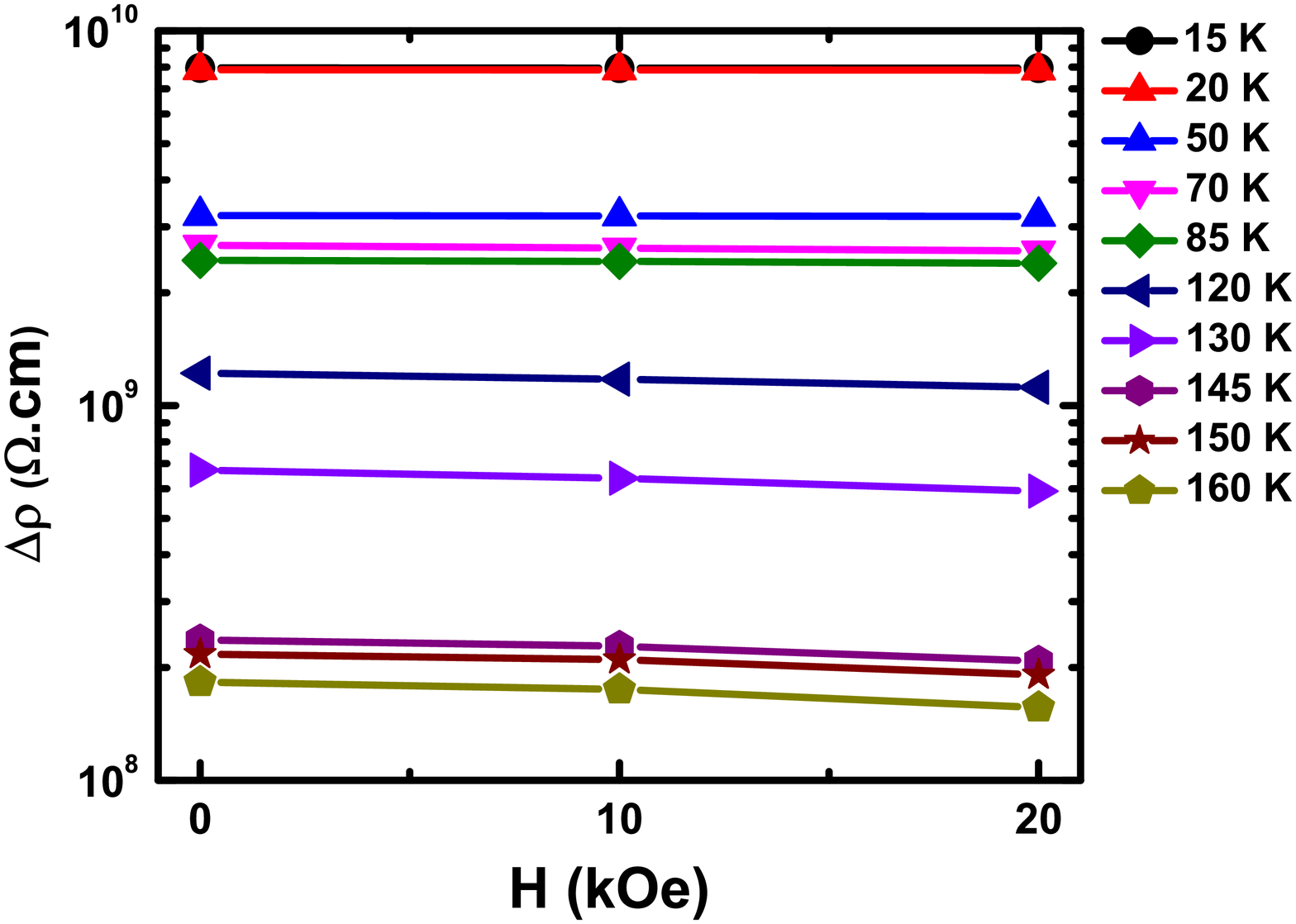}}
   \end{center}
\caption{Variation of the (a) $E_{th}$ (solid symbols), $\Delta E_{th}$ (open symbols), and (b) $\Delta\rho$ with the applied magnetic field ($H$). At low temperature, the dependence is extremely weak while with the rise in temperature it becomes progressively stronger.}

\end{figure}

\begin{figure}[ht!]
\centering
{\includegraphics[scale=0.25]{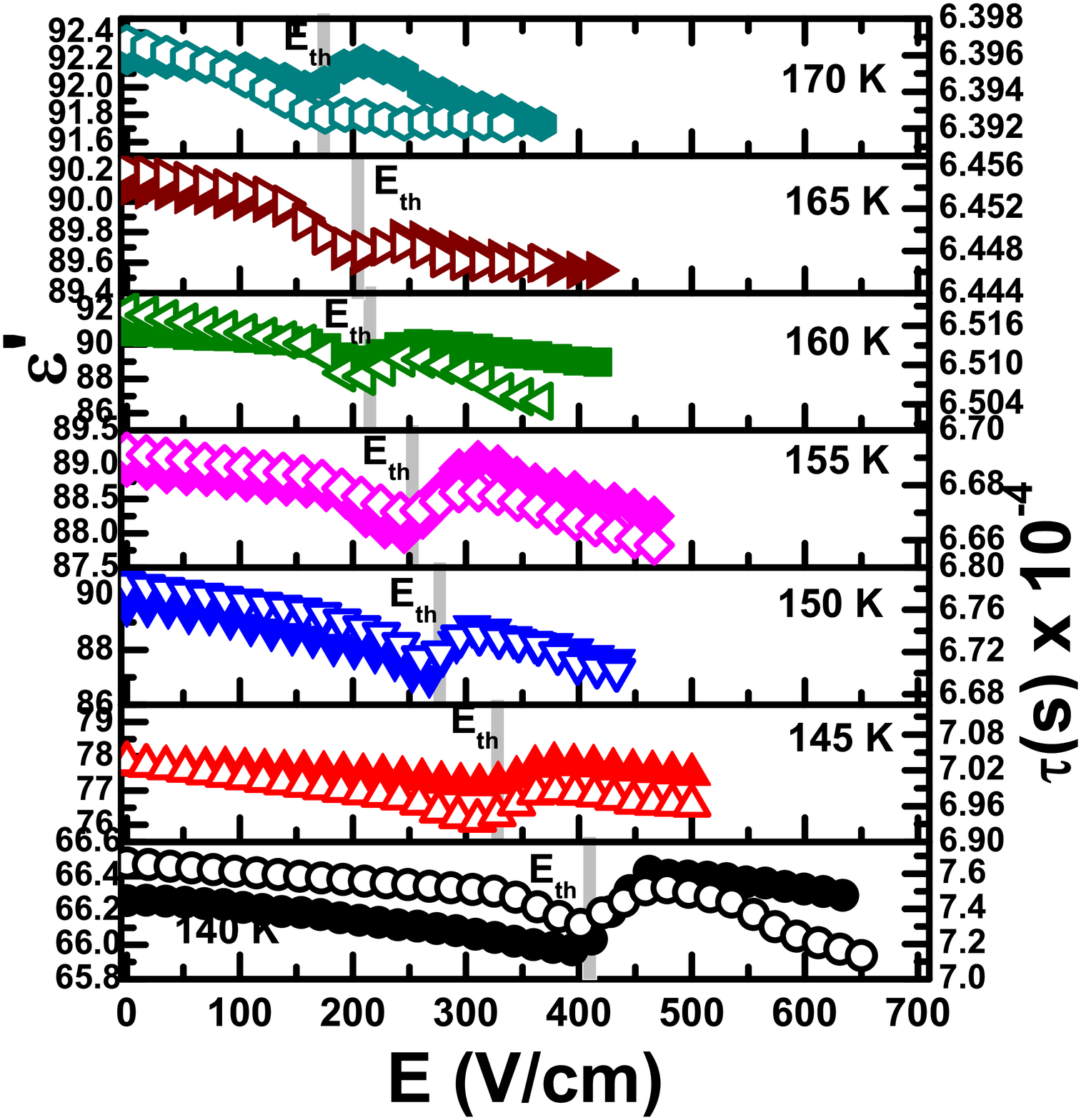}}
\caption{Variation of dielectric constant and relaxation time constant with electric field; solid (open) symbols represent the dielectric constant (relaxation time constant); anomalous features around the threshold electric field $E_{th}$ indicate influence of electric field driven transition on dielectric properties.}

\end{figure}

\begin{figure}[ht!]
\begin{center}
   \subfigure[]{\includegraphics[scale=0.25]{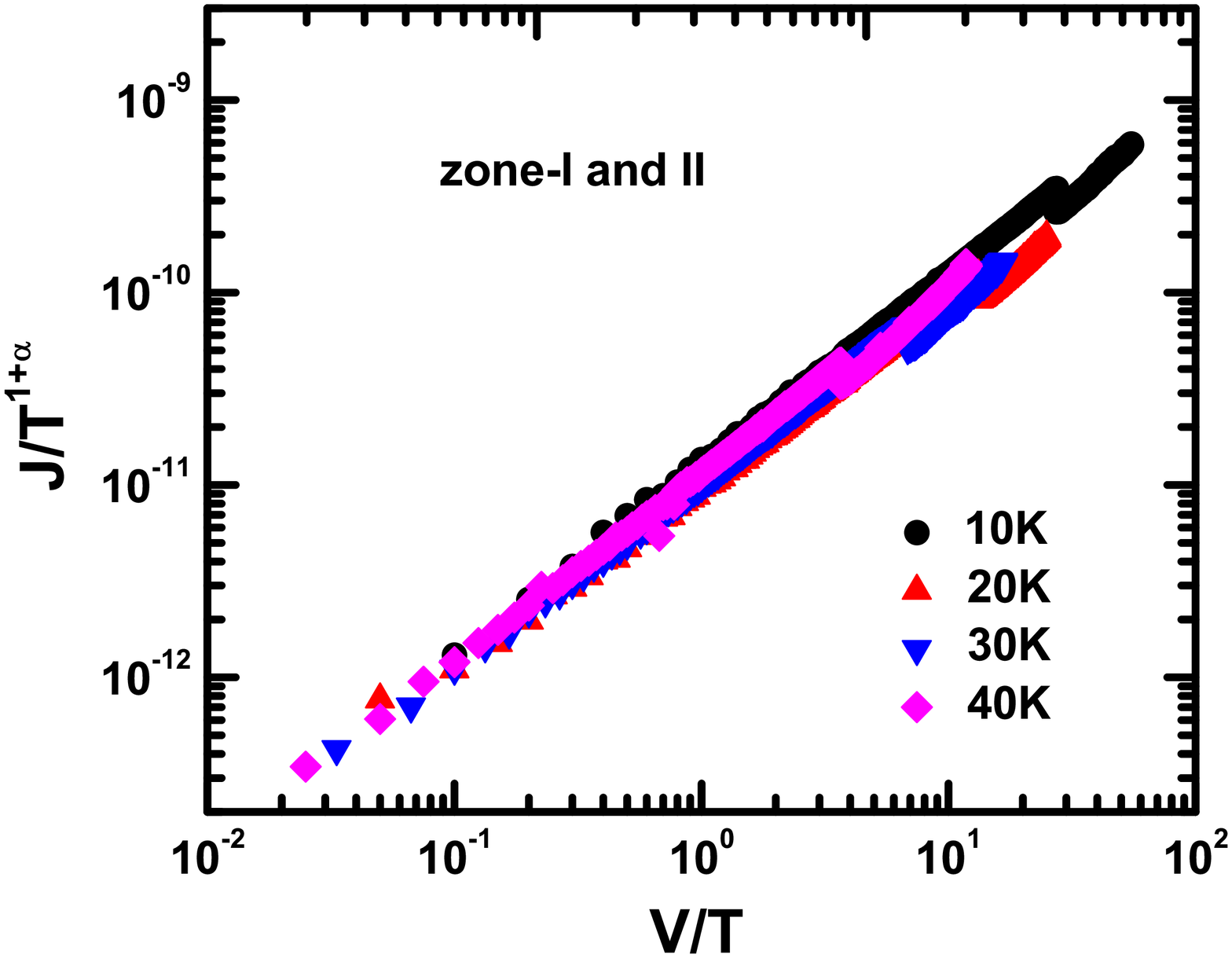}} 
   \subfigure[]{\includegraphics[scale=0.25]{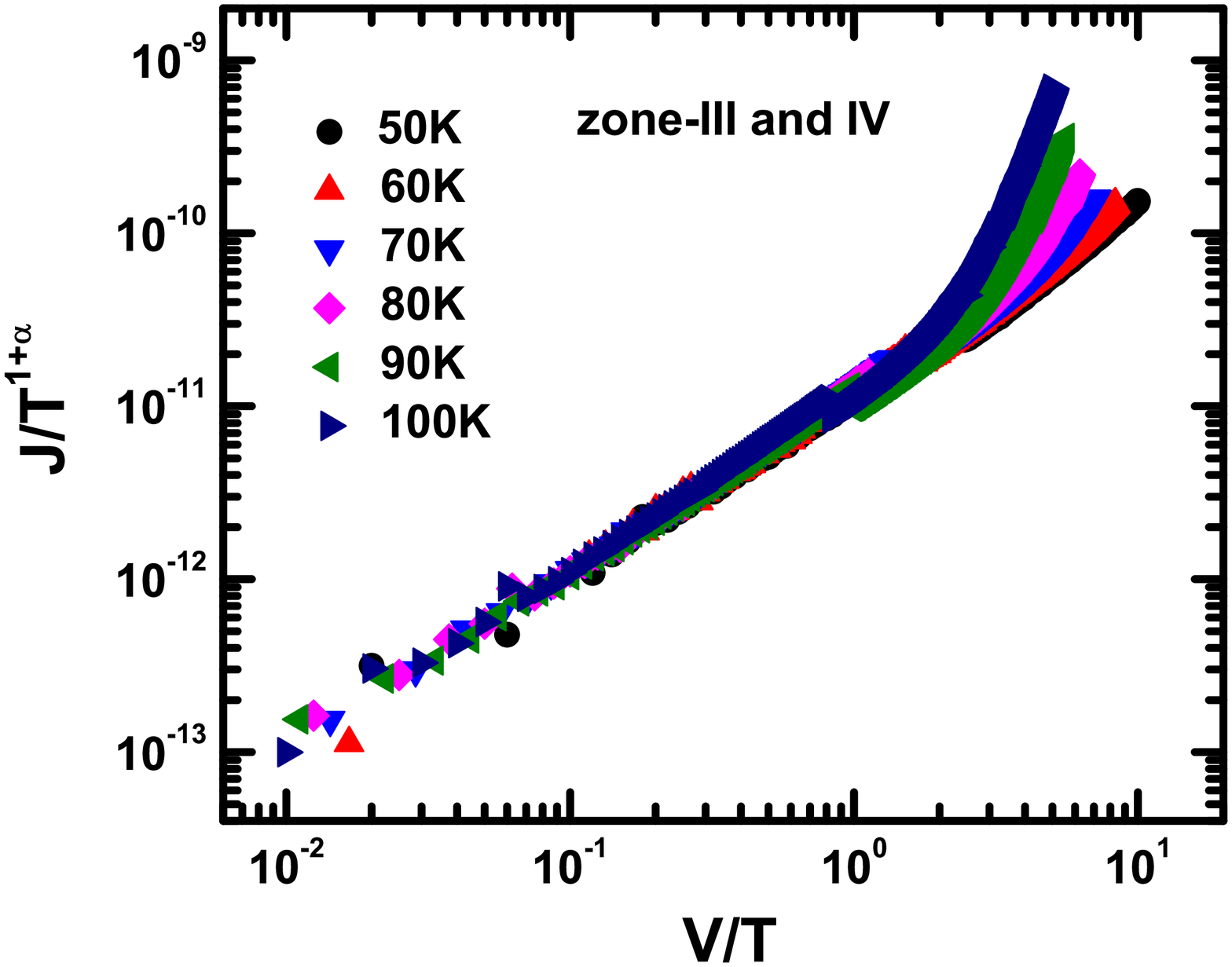}}
   \subfigure[]{\includegraphics[scale=0.25]{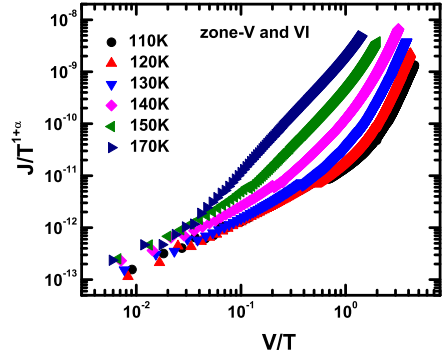}}
   \end{center}
\caption{The universal scaling, as per the model based on effective temperature of the charge carriers with respect to the lattice and heat balance between heat supplied by Joule heating and absorbed by charge carrier and lattice, appears to explain the $\rho(E,T)$ in the (a) zone I, II, and (b) III but (c) not in other zones. }
\end{figure}

\begin{figure*}[ht!]
\centering
{\includegraphics[scale=0.40]{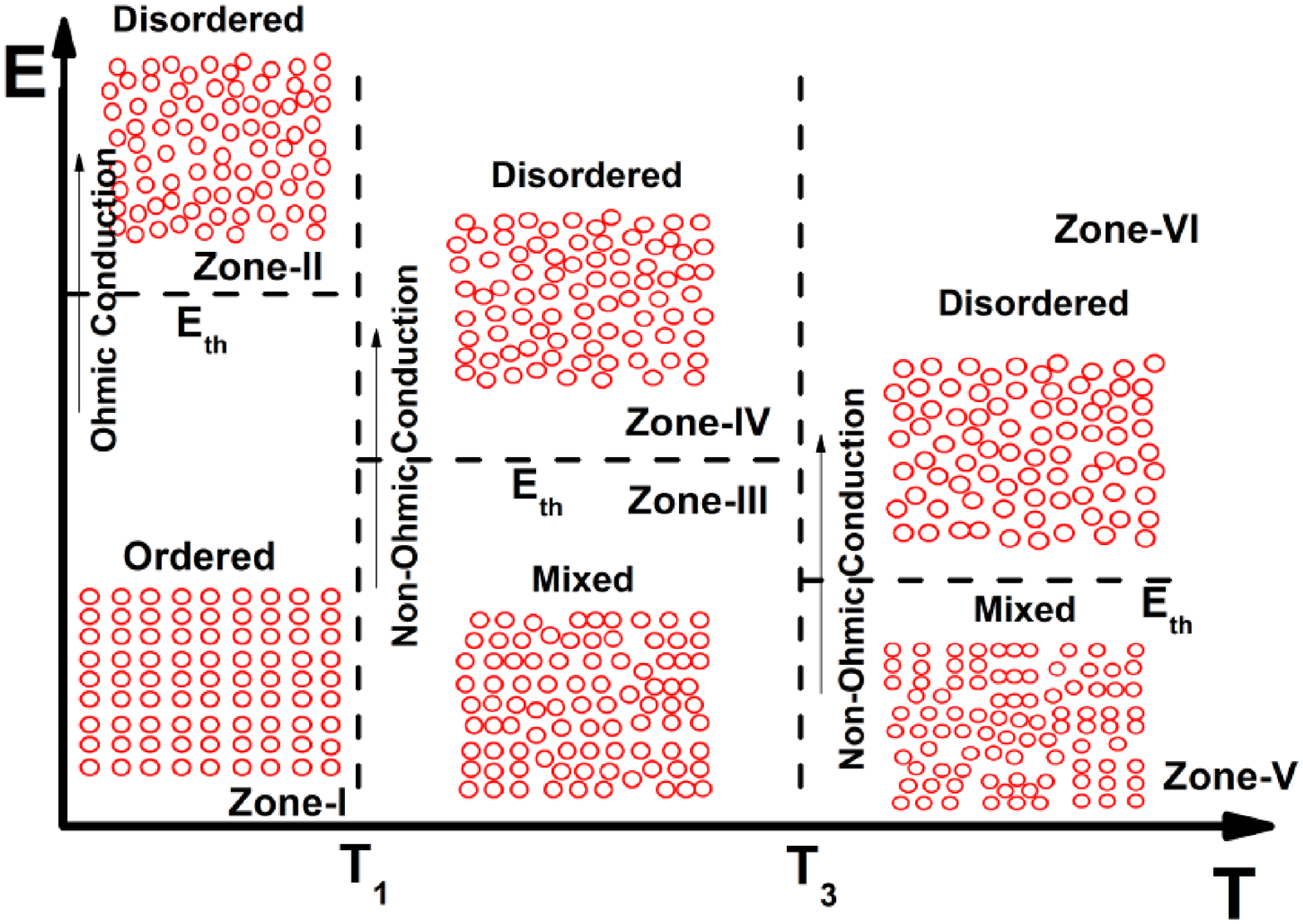}}
\caption{Schematic of the evolution of $E$- and $T$-driven charge conduction scenarios as a result of electromigration and Joule heating driven structural transitions in the point defect network. }
\end{figure*}

We concentrate here more on the observed electric-field-driven resistive switching features. Before we proceed further we point out that the overall conductivity $\sigma$ of the composite depends on the individual conductivities of W- and Z-type hexaferrites, $\sigma_W$ and $\sigma_Z$, respectively. Since, $\sigma_W$ and $\sigma_Z$ are comparable \cite{Ouyang} and hence this composite is not close to the percolation threshold, effective medium theory could be applicable here where the overall conductivity is given by $\sigma$ = $v_W\sigma_W$+$(1.0-v_W)\sigma_Z$, $v_W$ is the volume fraction of the W-type hexaferrite; $V_W$ = 0.8. Preliminary examination of the $\rho-E$ patterns shows that at low temperature (10-25 K), the LRS as well as HRS are essentially Ohmic with nearly field-independent resistivity. With further rise in temperature beyond 25 K, $\rho(E)$ turns out to be non-Ohmic and decreasing with the increase in $E$ both in LRS and HRS. We examine the $J-E$ and $\rho-E$ patterns - corresponding to both LRS and HRS - observed in different temperature regimes under zero magnetic field. In Fig. 4, we show the representative plots. Both at LRS and HRS, as the temperature is raised from 10 K, Ohmic transport crosses over to non-Ohmic one. Of course, prevalence of any interface and/or electrode controlled effects such as Fowler-Nordheim tunneling, Schottkey or Poole-Frenkel emission etc could be ruled out. In the cases where $E$-driven non-Ohmic transport is governed by Schottkey emission, $lnJ$ should follow $\propto E^{0.5}$ relation. In the present case, such a relation does not appear to be holding good. 

The $\rho-E$ patterns in the non-Ohmic conduction regime, evolve as the temperature is raised to cross over from phase-I to II, III, and IV. The phase-II and III do not exhibit any clear functional dependence of $\rho(E)$. Different functional forms from simple Arrhenius $\rho(E) = \rho_0.exp(E_0/E)$ to variable range hopping $\rho(E) = \rho_0.exp[(E_0/E)^{1/N}]$ ($N = 1+D$, $D$ = 2 or 3 for two or three-dimensional conduction) were found to be inadequate to describe the patterns satisfactorily as they do not provide unique best fitting \cite{Ma}. In phase-IV, $\rho(E)$ is found to follow three-dimensional VRH (3D-VRH) and power-law relation $\frac{\rho}{\rho_0}$ $\approx$ $(\frac{E_0}{E})^m$ with the exponent $m$ varying within $\sim$0.6-0.7 and $\sim$0.6-0.8 at LRS and HRS, respectively, more clearly. This evolution of noncompliance with any particular model to clear power law dependence of $\rho$ on $E$ signifies prevalence of a different conduction mechanism - in the case of $E$-driven one - from activated polaron hopping within the near-neighbor sites (manifested by exponential dependence of $\rho$ on $E$). It also indicates that this regime is in the vicinity of $E$-driven metal-insulator transition with effective diffusion length ($d_E$) smaller than the localization length ($\zeta_E$) \cite{Thouless}. Interestingly, as mentioned above, the $T$-driven conduction under a constant $E$ turns out to be governed by activated small polaron hopping with $E$- and $H$-dependent $U$ in all these regimes. The LRS to HRS transition under the sweeping applied $E$ is also associated with hysteresis (Fig. 2). The extent of hysteresis ($\Delta E_{th}$) decreases with the rise in temperature (Fig. 3c).    

The hysteresis in the $J-E$ or $\rho-E$ plots observed around $E_{th}$ could result from Joule heating \cite{Fursina}. At LRS, the heating is large, as large current flows through the sample. This leads to upward shift of $E_{th}$ in the forward branch of the $\rho-E$ pattern (Fig. 2) since Joule heating extends the persistence of LRS at an $E(T)$ $>$ $E_{th}(T)$. Along the reverse branch, because of HRS, Joule heating is smaller and hence it persists down to $E_{th_{expected}}(T)$. Difference between actual and expected $E_{th}$ gives rise to the observed hysteresis. Obviously, as the operating temperature enhances, the extent of resistive change ($\Delta\rho$) decreases and, consequently, the difference in the extent of Joule heating between LRS and HRS also decreases. The hysteresis too drops down as a result. In order to examine the influence of Joule heating more specifically, we have scanned the current-voltage characteristics at $\sim$10 K over different time spans. The time scale dependence of the $E_{th}$ - observed during forward and reverse scanning - is shown in the supplementary document \cite{supplementary}. Indeed, the $E_{th}$ obtained from forward scan enhances with the time scale while that noted during the reverse scan shows relatively weaker time scale dependence.  

Application of magnetic field $H$ ($\sim$10-20 kOe) does not have any significant influence below $T_1$. However, at higher temperature, reasonably substantial magnetoresistance $\frac{\Delta\rho(H)}{\rho(0)}$ could be observed. The decrease in $E_{th}$ under $H$ too is noticeable at higher temperature while the dependence of $U$ on $H$ turns out to be discernible only in phase-IV. This could be because of rather stable magnetic structure with large magnetocrystalline anisotropy at lower temperature. With the rise in temperature, the enhancement of thermal energy induces switching of magnetization under even moderate magnetic field ($\sim$10-20 kOe) and this, in turn, possibly yields stronger influence of magnetic field on the electrical properties in this temperature range. Of course, no magnetic transition could be observed around $T_1$, $T_2$, or $T_3$. Therefore, change in magnetic structure cannot play any role here. The dependence of $E_{th}$, $\Delta E_{th}$, and $\Delta\rho$ on $H$ is shown in Fig. 5. More data on magnetoresistance are available in the supplementary document \cite{supplementary}. The exponent $m$ which describes the $\rho-E$ patterns under different $H$ does not, however, exhibit any significant dependence on $H$.

The resistive transition has been probed further by dielectric spectroscopy recorded under finite dc bias in order to examine whether the transition is thermodynamic. It shows anomalous feature in dielectric constant ($\epsilon$) at the $E_{th}(T)$. The complex plane impedance spectra - recorded across 100 Hz to 5 MHz - have been fitted by an equivalent circuit model comprising of resistance and capacitance for the intrinsic and sample-electrode interface dielectric response. The fitting are shown in the supplementary document \cite{supplementary}. Variation of the intrinsic dielectric constant ($\epsilon$) and the relaxation time constant ($\tau_{in}$ = $R_{in}C_{in}$; $C_{in}$ is intrinsic capacitance) with $E$ is shown in Fig. 6. Change in the conduction dynamics across LRS to HRS appears to have significant influence on the dielectric response. The dielectric constant ($\epsilon$) as well as the relaxation time constant ($\tau$) exhibit clear anomalous features around the $E_{th}$. This observation points out that the electric-field-driven resistive transition is indeed a thermodynamic one. Of course, because of the limitation of experimental set-up, dc bias higher than 40 V could not be applied and, hence, corresponding dielectric data for the low temperature regime - where scanning across $E_{th}$ requires application of dc bias far greater than 40 V - could not be generated.     

We also examine the applicability of the universal scaling model \cite{Abdalla} in the present case. This model accounts for $E$ and $T$-driven conductivity in terms of effective charge carrier temperature $T_{eff}$ which is obtained from the heat balance between Joule heating of the system and energy dissipation through the lattice. At the higher $E$ end, charge carriers gain energy and become `hot'. Under such circumstances, charge conduction primarily follows variable range hopping mechanism driven by $T_{eff}$. In the Fig. 7, we show the scaling of the conductivity data in the present case - plotted as $J/T^{1+\alpha}$ versus $V/T$ where $J$ and $V$ are current and voltage respectively. The scaling considers power-law temperature dependence of conductivity $\sigma$ $\propto$ $T^{\alpha}$ and non-Ohmic conduction with $J$ $\propto$ $V^{\beta}$, $\beta$ = $1+\alpha$. Clearly, data corresponding to the zone I, II, and III appear to follow the scaling for $\alpha$ = 0.5 while those corresponding to the zone IV, V, and VI exhibit departure (zones have been marked in the $E-T$ phase diagram shown in Fig. 8). In these latter zones, the concept of $T_{eff}$ does not hold good. 

Based on the observations and analyses described above, a qualitative picture is being attempted to develop here for describing the evolution of the charge conduction and resistive transition across the entire $E-T$ regime (Fig. 8). It relies on influence of structural evolution in the point-defect network on the electronic conductivity. The charge conduction is considered to be, primarily, electronic. In spite of electromigration of defects, ionic conduction is not so significant in comparison to the electronic one. We did not attempt here to determine the electronic and ionic conduction separately. As described earlier, this system contains intrinsic inhomogeneity. The EDX mapping of the elemental concentration provides detailed picture of the inhomogeneity. The concentration of the point defects such as cation vacancies and oxygen excess - as determined from the quantitative analysis of the EDX data - is also quite substantial \cite{supplementary}. This point defect network is expected to influence the charge conduction significantly. The $E$ and $T$-driven resistive transitions signify, as shown in Fig. 8, emergence of different phases - from ordered to mixed order/disorder or to granular or disordered structures within the point defect network. Justification for the emergence of such structures could be obtained from the analyses of the $\rho(E,T)$ patterns in different zones of the phase diagram shown in Fig. 8. 

Within zone-I (i.e., below $T_1$ and corresponding $E_{th}$), the defect network could be ordered (Fig. 8). The $E$-driven conduction turns out to be Ohmic in this regime while the $T$-driven conduction follows activated small polaron hopping mechanism as described earlier. Electromigration of point defects under higher electric field, in this situation, leads to defect order to disorder transition and is instrumental in inducing the $E$-driven resistive transition. Of course, migration of defects could also be influenced by Joule heating in the LRS where the amount of current flow through the system is large. The $\rho(E,T)$ patterns collapse into universal scaling law (Fig. 7a) with $\alpha$ = 0.5. Deviation from the universal scaling in zone-II is detectable where Ohmic conduction exhibits a sharp drop at $E_{th}$ because of enormously enhanced scattering from the `disordered' defects. As the temperature is raised beyond $T_1$, the `defect ordered' state possibly undergoes $T$-driven transition to a mixed state of coexisting ordered and disordered states (zone-III in Fig. 8). The volume fraction of the ordered state could be higher. The Ohmic conduction gives way to a conduction with complex functional $E$ dependence which, in turn, is abruptly restricted (akin to disorder driven metal-insulator transition) in the `defect disordered' state at $E_{th}$ (zone-IV, Fig. 8) because of confinement of the charge carriers. Interestingly, in zone-III too, $\rho(E,T)$ exhibits universal scaling with $\alpha$ = 0.5 (Fig. 7b). However, deviation from universal scaling in zone-IV is clearly visible. As the temperature is raised even further, the $T$-driven transition in the defect network yields finer domain or mixed state structure with higher volumetric proportion of disordered state (Fig. 8). The conduction exhibits clear compliance with either three-dimensional VRH (3D-VRH) or power law dependence on $E$ in zone-V and VI. Of course, departure from universal scaling is also clearly visible in these regimes (Fig. 7c). Since no clear distinction in the applicability of universal scaling model could be noticed for the data corresponding to phase-II and phase-III, we ignore this distinction while discussing the evolution of the point defect structure. 

The $T$-driven conduction, throughout these regimes, follows activated small polaron hopping model with decrease in $U$ with the rise in $E$. The $E$-dependence of $U$ results from electromigration driven local change in the structure which, in turn, could give rise to decrease in the activation barrier $U$. Similar structural-change-driven decrease in $U$ with the increase in $E$ has earlier been observed in other transition metal oxide systems \cite{Kuang}. Application of $H$ up to 20 kOe turns out to be influencing both the $U$ as well as all the relevant parameters of $E$-driven conduction at higher temperature. Notably, $\rho(E,T)$, in both the zone-V and VI, deviate significantly from the universal scaling law. Electric field and temperature driven order-disorder transition in point defect network has earlier been observed in different compounds \cite{Tomura,Wang,Poloni,Lim}. Electric-field-induced generation of charge-neutral anti-Frenkel defect pairs have recently been reported \cite{Evans} in transition metal oxide Eu(Ti,Mn)O$_3$ for controlling the electronic conduction without influencing the structural, magnetic, and electric properties. Electromigration has also been observed at low temperature in different metallic and non-metallic systems \cite{Rhoden,Xiang}. Using scanning tunneling electron microscopy (STEM) and electron energy loss spectroscopy (EELS), the defect network and its transition could be observed in different cases \cite{Tomura,Evans}. The charge conduction changes dramatically as a result of such transitions in the structure of defect network. 

In the present case, compliance/non-compliance with universal scaling along with applicability of models such as activated small polaron hopping, 3D-VRH, and power law dependence help to construct the defect phases shown in Fig. 8. Ordered structure offers Ohmic conduction and compliance with small polaron hopping model. It also yields scalability to universal scaling model. Disordered structure, on the other hand, indicates deviation from universal scaling even though $E$-driven conduction remains Ohmic in this range. Mixed structure, especially when ordered network starts acquiring a certain extent of disorder, deviation from Ohmic conductivity starts becoming visible. Even then, within this range, universal scaling appears to be applicable. This could be because of the fact that the entire electron-lattice system could reach a thermal equilibrium with effective temperatures $T_{eff}$ and $T_{latt}$ for both the charge carriers and the lattice. In the disordered structure, such equilibrium is not reached. As a result, clear deviation from universal scaling could be observed in zone-IV, V, and VI. It is important to mention here that the compliance with universal scaling law could only be observed for $\alpha$ = 0.5. This is different from the Thouless model which predicts $\alpha$ = 1.0. Of course, deviation from Thouless model has been noticed ($\alpha$ $\approx$ 4.0-6.0) in different conducting polymer systems \cite{Park}. 

It is also important to mention that in all the previous studies of $E$-driven charge conduction and resistive transition in a score of different transition metal oxide systems \cite{Budhani,Joung,Lim}, either Ohmic or activated hopping or variable range hopping conduction has been noticed. Power law dependence on $E$ has not been reported. Applicability of 3D-VRH and power law $E$ dependence indicates influence of `localization' as well. Although, this power-law dependence of conductivity was proposed originally for two-dimensional conduction, it was shown later \cite{Imry} to be valid for three-dimensional systems as well subject to the fulfilment of the condition of three-dimensional resistivity being less than a critical one to extend the localization length beyond the sample thickness. Within the $E$ and $T$ range corresponding to the zone-V and VI, the localization length ($\zeta_E$) should be comparable or greater than the charge diffusion length ($d_E$ is the length scale across which the charge carriers diffuse during $\tau_E$) \cite{Imry}. We further mention, in this context, that in the ferroelectric compounds, domain wall conductivity assumes different models including Mott or Efros-Shklovskii \cite{Scott}. The charge conduction via domain wall channels leads to rise in conductivity at low temperature. This is in sharp contrast to the observations made here. Therefore, domain wall conductivity is not quite relevant in the present case. The conduction as well as the defect order to disorder transition observed here are the bulk effects. It is further supported by the observation of anomalous jump in the intrinsic $\epsilon$ and $\tau$ at the characteristic $E_{th}$.      

\begin{figure}[ht!]
\centering
{\includegraphics[scale=0.25]{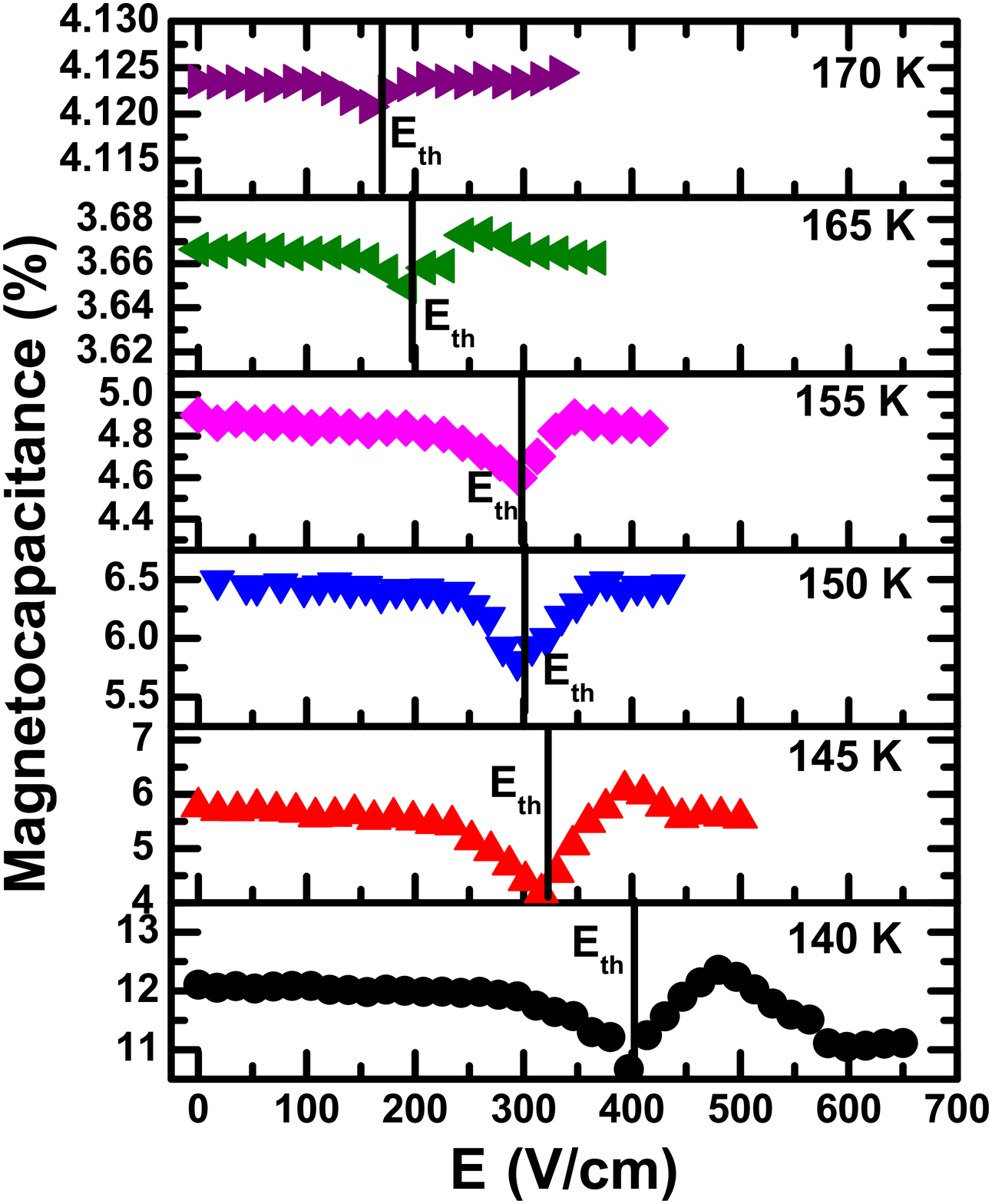}}
\caption{Variation of the magnetocapacitance with electric field across the resistive transition at $E_{th}$ at different temperatures. Anomalous dip in the magnetocapacitance around $E_{th}$ is clearly visible. Yet, no significant change in the field dependence of magnetocapacitance between LRS and HRS is noticeable. }
\end{figure}

Finally, we examined whether the $E$-driven transition from LRS to HRS influences the magnetoelectric effect. In ferroelectric HfO$_2$, sequential ferroelectric-resistive-ferroelectric switching was observed by others \cite{Max} and the role of the oxygen vacancies examined. Influence of ferroelectric switching on resistive one and vice versa was noted. In the present case, we carried out the dielectric spectroscopy measurements under $\sim$10 and $\sim$20 kOe fields. In Fig. 9, the extracted capacitance ($C$) and magnetocapacitance [\{$C(0)-C(H)/C(0)$\}$\times$100] (under $\sim$20 kOe field) are shown across the respective $E_{th}(T)$. The magnetocapacitance under both $\sim$10 and $\sim$20 kOe fields are shown in the supplementary document \cite{supplementary}. Large magnetocapacitance was earlier shown to be signifying strong coupling between ferroelectric and magnetic order parameters (i.e., multiferroicity) in, for example, BiMnO$_3$ \cite{Kimura} and LuFe$_2$O$_4$ \cite{Subramanian}. In the present case, the magnetocapacitance is found to be substantial ($\sim$4\%-12\%) both at LRS and HRS. Anomaly in the magnetocapacitance at $E_{th}(T)$ is clearly visible. However, no radical change in the magnitude (i.e., significant jump in the magnitude of the magnetocapacitance) and $E$ dependence of magnetocapacitance could be observed as a result of LRS to HRS transition. Therefore, the electric field driven transition in the defect structure does influence the intrinsic coupling between polarization and magnetization by a certain extent. The moderate change in the magnetocapacitance could be traced to the moderate change in the $\Delta\rho/\rho$. Detailed investigation of the crystallographic and electronic structures of the defect phases (corresponding to LRS and HRS) needs to be carried out seperately to unravel the origin of the moderate change in the magnetocapacitive effect. This will be done in near future.    

\section{Summary}
In summary, we observed an interesting electric-field-driven resistive transition in an off-stoichiometric composite of W- and Z-type multiferroic compound SrCo$_2$Fe$_{16}$O$_{27}$/Sr$_3$Co$_2$Fe$_{24}$O$_{41}$. The dielectric constant as well as relaxation time constant too exhibit anomalous jump near the threshold field. This system also exhibits several temperature-driven transitions in the 10-200 K range and these transitions coincide with the transitions probed by resistivity and dielectric data. The electric field driven charge conduction evolves from Ohmic to non-Ohmic across these transitions both in the low and high resistive states. The complete temperature and electric field driven phase transitions has been mapped across the 10-200 K and 0-6000 V/cm regimes. Compliance and non-compliance with universal scaling law together with the resistive jump helps in formulating a qualitative picture of charge conduction based on electric field and temperature driven structural evolution of point defect network. Such a network forms in this off-stoichiometric composite due to the presence of cation vacancies and/or excess oxygen. Quantitative mapping of the elemental concentration across several regions of the microstructure - obtained from energy dispersive x-ray spectra - provides information on local variation of the compositional homogeneity and thus offers the basis for formulating the qualitative picture on formation of point defect network. The substantial magnetocapacitive effect ($\sim$4\%-12\%) reflects clear influence of the electric field driven transition. Yet the change in the magnitude of the magnetocapacitive effect, across the electric field driven transition in the point defect network, is moderate. These observations provide insight about the temperature, electric, and magnetic field driven charge conduction in these off-stoichiometric composite systems and their correlation with multiferroicity. Importantly, the variation in the extent of non-stoichiometry is found to influence the resistive transition and charge conduction features significantly. Therefore, both electric field driven resistive jumps and the multiferroicity could be tuned to pave the way for electric and magnetic field sensor applications. 

\begin{center}
$\textbf{ACKNOWLEDGMENTS}$
\end{center}
Two of the authors (S.M., first author and A.S.) acknowledge support from DST-INSPIRE fellowship of Government of India during the work.

\end{document}